\documentclass[twocolumn, 10pt]{IEEEtran}
\usepackage{hyperref} 
\usepackage{graphicx}
\usepackage{amssymb}
\usepackage{amsmath}
\usepackage{mathtools}
\usepackage{cite}
\usepackage{stfloats}
\usepackage{subfigure}
\usepackage{epstopdf}
\usepackage{psfrag}
\usepackage[mathscr]{euscript}
\usepackage{acronym}  
\usepackage{booktabs} 
\usepackage[table]{xcolor}
\usepackage{slashbox,pict2e}

\acrodef{CCDF}{complementary cumulative distribution function}
\acrodef{CF}{characteristic function}
\acrodef{PPP}{Poisson point process}
\acrodef{CSI}{channel state information}
\acrodef{OFDM}{orthogonal frequency division multiplexing}
\acrodef{OFDMA}{orthogonal frequency division multiple access}

\acrodef{RV}{random variable}
\acrodef{i.i.d.}{independent, identically distributed}
\acrodef{PMF}{probability mass function}
\acrodef{PDF}{probability distribution function}
\acrodef{PGFL}{probability generating functional}
\acrodef{CDF}{cumulative distribution function}
\acrodef{ch.f.}{characteristic function}
\acrodef{AWGN}{additive white Gaussian noise}
\acrodef{SNR}{signal-to-noise ratio}
\acrodef{LRT}{likelihood ratio test}
\acrodef{DRT}{distance ratio test}
\acrodef{GLRT}{generalized likelihood ratio test}
\acrodef{CRLB}{Cram\'{e}r-Rao lower bound}
\acrodef{CRB}{Cram\'{e}r-Rao bound}
\acrodef{ZZLB}{Ziv-Zakai lower bound}
\acrodef{ZZB}{Ziv-Zakai bound}
\acrodef{LoS}{line-of-sight}
\acrodef{ToF}{time-of-flight}
\acrodef{NLoS}{non-line-of-sight}
\acrodef{GDOP}{geometric dilution of precision}
\acrodef{GPS}{Global Positioning System}
\acrodef{FIM}{Fisher information matrix}
\acrodef{PEB}{position error bound}
\acrodef{SPEB}{squared position error bound}
\acrodef{TOA}{time-of-arrival}
\acrodef{TOF}{time-of-flight}
\acrodef{WSN}{wireless sensor network}
\acrodef{MAC}{medium access control}
\acrodef{RSS}{received signal strength}
\acrodef{WAF}{wall attenuation factor}
\acrodef{TDOA}{time difference-of-arrival}
\acrodef{RF}{radiofrequency}
\acrodef{RTT}{round-trip time}
\acrodef{AOA}{angle-of-arrival}
\acrodef{MF}{matched filter}
\acrodef{ED}{energy detector}
\acrodef{ML}{maximum likelihood}
\acrodef{MSE}{mean-square error}
\acrodef{RMSE}{root-mean-square error}
\acrodef{LEO}{localization error outage}
\acrodef{ppm}{part-per-million}
\acrodef{ACK}{acknowledge}
\acrodef{UWB}{Ultrawide bandwidth}
\acrodef{TNR}{threshold-to-noise ratio}
\acrodef{LS}{least squares}
\acrodef{IR-UWB}{impulse radio UWB}
\acrodef{FCC}{Federal Communications Commission}
\acrodef{TH}{time-hopping}
\acrodef{PPM}{pulse position modulation}
\acrodef{MUI}{multi-user interference}
\acrodef{PDP}{power delay profile}
\acrodef{BPZF}{band-pass zonal filter}
\acrodef{SIR}{signal-to-interference ratio}
\acrodef{RFID}{radio frequency identification}
\acrodef{WPAN}{wireless personal area network}
\acrodef{WWB}{Weiss-Weinstein bound}
\acrodef{DP}{direct path}
\acrodef{MF}{matched filter}
\acrodef{MMSE}{minimum-mean-square-error}
\acrodef{SBS}{serial backward search}
\acrodef{SBSMC}{serial backward search for multiple clusters}
\acrodef{NBI}{narrowband interference}
\acrodef{WBI}{wideband interference}
\acrodef{INR}{interference-to-noise ratio}
\acrodef{CR}{channel response}
\acrodef{CIR}{channel impulse response}
\acrodef{CR}{channel  response}
\acrodef{RADAR}{radar}
\acrodef{MUR}{Multistatic radar}
\acrodef{JBSF}{jump back and search forward}
\acrodef{HDSA}{high-definition situation-aware}
\acrodef{RRC}{root raised cosine}
\acrodef{ST}{simple thresholding}
\acrodef{BTB}{Bellini-Tartara bound}
\acrodef{P-Max}{$P$-Max}  
\acrodef{MIMO}{multiple-input multiple-output}
\acrodef{MAP}{maximum a posteriori}
\acrodef{FG}{factor graph}
\acrodef{OP}{outage probability}
\acrodef{WED}{wall extra delay}
\acrodef{RMS}{root mean square}
\acrodef{SPAWN}{sum-product algorithm over a wireless network}
\acrodef{MDD}{minimum distance distribution}
\acrodef{MAP}{maximum a posteriori probability}
\acrodef{PAR}{probabilistic association rule}
\acrodef{AP}{access point}
\acrodef{HD}{half-duplex}
\acrodef{FD}{full-duplex}
\acrodef{IC}{interference cancellation}
\acrodef{HDHN}{hybrid-duplex heterogeneous network}
\acrodef{TDD}{time-division duplexing}
\acrodef{FDD}{frequency-division duplexing}
\acrodef{SINR}{signal-to-interference-plus-noise ratio}
\acrodef{UAV}{unmanned aerial vehicle}
\acrodef{GCS}{ground control station}
\acrodef{LTE}{long term evolution}
\acrodef{VANET}{vehicular ad hoc network}
\acrodef{3GPP}{3rd Generation Partnership Project}
\usepackage{color}
\usepackage{dsfont}
\usepackage{bbm}

\newcommand{\PL}[1]{\alpha_{#1}} 

\newcommand{\Pm}{P_\text{m}}
\newcommand{\PI}{P_\text{I}}

\newcommand{\Lm}{\ell_\text{m}}
\newcommand{\Lma}{\ell_\text{m}^{-\alpha_\text{m}(\ell_\text{m})}}
\newcommand{\Li}{\ell_\text{I}}
\newcommand{\Lia}{\ell_\text{I}^{-\alpha_\text{I}(\ell_\text{I})}}
\newcommand{\Hm}{h_\text{m}}
\newcommand{\Hi}{h_\text{I}}
\newcommand{\HL}{\overline{H_\text{L}}}
\newcommand{\HN}{\overline{H_\text{N}}}
\newcommand{\Beta}[1]{\beta_\text{#1}(\ell_\text{#1})}

\newcommand{\Km}{K_\text{m}}
\newcommand{\Ki}{K_\text{I}}

\newcommand{\Ws}[2]{{W_{}^{}}} 

\newcommand{\Dist}[2]{d_\text{#1}^{(\text{#2})}}

\newcommand{\TSIR}[2]{{\tau_{}^{}}}

\newcommand{\Pout}{p_{\text{o}}}

\newcommand{\Pouth}{\hat{p}_\text{o}}

\newcommand{\PLoS}{p_{\text{L}}(\ell_i)}
\newcommand{\PNLoS}{p_{\text{N}}(\ell_i)}
\newcommand{\PLoSm}{p_{\text{L}}(\ell_\text{m})}
\newcommand{\PLoSi}{p_{\text{L}}(\ell_\text{I})}
\newcommand{\PNLoSm}{p_{\text{N}}(\ell_\text{m})}
\newcommand{\PNLoSi}{p_{\text{N}}(\ell_\text{I})}

\DeclareMathAlphabet{\mathsf}{OML}{cmbr}{m}{it}

\newtheorem{theorem}{Theorem}
\newtheorem{lemma}{Lemma}
\newtheorem{corollary}{Corollary}

\newcommand{\bd}{\begin{description}}
\newcommand{\ed}{\end{description}}
\newcommand{\be}{\begin{enumerate}}
\newcommand{\ee}{\end{enumerate}}
\newcommand{\bi}{\begin{itemize}}
\newcommand{\ei}{\end{itemize}}
\newcommand{\bl}{\begin{list}}
\newcommand{\el}{\end{list}}
\newcommand{\bt}{\begin{tabbing}}
\newcommand{\et}{\end{tabbing}}

\newcounter{eqncnt}

\newcounter{eqnback}

\acrodef{BS}{base station}
\acrodef{G2A}{ground-to-air}
\acrodef{A2G}{air-to-ground}
\acrodef{A2A}{air-to-air}
\acrodef{G2G}{ground-to-ground}
\acrodef{IoT}{internet of things}

\begin{document}

\newcommand{\paperTitle}{{Impact of an Interfering Node\\ on Unmanned Aerial Vehicle Communications}}
 
 

\title{\paperTitle}

\author{
	\IEEEauthorblockN{
		Minsu Kim and 
		Jemin~Lee, \textit{Member, IEEE} 
	}\\[0.5em]
	%
	    \thanks{
	       M.\ Kim and J.\ Lee are with the Department of Information and
	       Communication Engineering, Daegu Gyeongbuk Institute of Science and
	       Technology (DGIST), Daegu 42988, South Korea
	       (e-mail: \texttt{ads5577@dgist.ac.kr}, \texttt{jmnlee@dgist.ac.kr}).
	     }
	      \thanks{The material in this paper was presented, in part, at the Global Communications Conference, Abu Dhabi, UAE, Dec. 2018 \cite{KimLee:18}
	       }
	       \thanks{
	       The corresponding author is J. Lee. 
			}
}

\maketitle 

%

%

%
\setcounter{page}{1}
\acresetall
\begin{abstract}
Unlike terrestrial communications,
\ac{UAV} communications have some advantages such as the \ac{LoS} environment and flexible mobility.
However, the interference will be still inevitable. 
In this paper, 
we analyze the effect of an interfering node on the \ac{UAV} communications
by considering the \ac{LoS} probability and different channel fading for \ac{LoS} and \ac{NLoS} links, which are affected by {horizontal and vertical distances} of the communication link.
We then derive a closed-form outage probability in the presence of an interfering node for all the possible scenarios and environments of main and interference links.
After discussing the impacts of transmitting and interfering node parameters on the outage probability, 
we show the existence of the optimal height of the \ac{UAV} that minimizes the outage probability.
We also show the \ac{NLoS} environment can be better than the \ac{LoS} environment if the average received power of the interference is more dominant than that of the transmitting signal on \ac{UAV} communications.
Finally, we analyze the {network outage probability} for the case of multiple interfering nodes using stochastic geometry and the outage probability of the single interfering node case, and show the effect of the interfering node density on the optimal height of the \ac{UAV}.
\end{abstract}
%
\acresetall
%
%
\section{Introduction}\label{sec:Intro}
%
As the \ac{UAV} technology develops, reliable \ac{UAV} communications have become necessary.
However, since \ac{UAV} communications are different from conventional terrestrial communications, it is hard to apply the technologies used in terrestrial communications to \ac{UAV} communications\cite{GupJainVas:16,ZenZhaLim:16,HayYanMuz:16,KhaGuvMat:18}.
Especially, unlike terrestrial communications, \ac{UAV} communications can have \ac{LoS} environments between a \ac{UAV} and a ground device, and between \acp{UAV}.
When the main link is in the \ac{LoS} environment, the received main signal power will increase due to better channel fading and lower path loss exponent compared to the \ac{NLoS} environment.
%
It also means that in the presence of an interfering node, the interfering signal can be received with larger power as the interfering link can also be in the \ac{LoS} environment \cite{ZhaYan:17,ChoLiuLee:18}.

\ac{UAV} communications have been studied in the literature, mostly focused on the optimal positioning and trajectory of the \ac{UAV}.
The height of the \ac{UAV} affects the communication performance in different ways.
As the height increases,
the \ac{UAV} forms the \ac{LoS} link with higher probability, which is modeled by the \ac{LoS} probability in \cite{HouKanLar:14},
but the distance to the receiver at the ground increases as well.
By considering this relation, the optimal height of the \ac{UAV} in terms of the communication coverage in the \ac{A2G} channel is presented in \cite{HouKanLar:14,BorElYan:16,AlzKeyLag:17}.
For the case of using a \ac{UAV} as a relay, the optimal height and position of \acp{UAV} have also been presented in \cite{CheFenZhe:18,AzaRoChePol:18}.
The optimal deployment and trajectory of the \ac{UAV} have been presented to minimize the power consumption in \cite{MozSaaBenDeb:16,MozSaaBenDe:16}.
The height of the \ac{UAV} and the power allocation factor have been jointly optimized to minimize the hybrid outage probability in \cite{LiuLeeQuek:19}.
The \ac{UAV} trajectory and transmit power control have been jointly optimized to minimize the outage probability in \cite{ZhaZhaHe:18} and to maximize the average secrecy rate in \cite{ZhaWuCui:18}.
The work in \cite{LyuZenZha:16} jointly optimized the throughput and the access delay using a cyclical multiple access scheme,
and the work in \cite{ZhaZenZh:17} jointly optimized the communication time allocation and the \ac{UAV} trajectory to maximize spectrum efficiency and energy efficiency.
However, the works in \cite{ZhaWuCui:18,ZhaZhaHe:18,LyuZenZha:16,ZhaZenZh:17} did not consider the \ac{LoS} probability,
and all of those works analyzed and optimized for the \ac{UAV} communications in the absence of an interfering node.
Since the interference is an inevitable factor in the current and future networks, the impact of the interference on the \ac{UAV} communications needs to be investigated carefully.

%
Recently, the interference has been considered in some works such as \cite{MozSaaBenDeb:15,MozSaaBen:16,CheDhi:17,AlzYan:18,WanZhaTia:18,CheAlzYan:19,KimLeeQue:19,AzaRosPol:19,AzaGerGar:19,WuZenZha:18,LeeEomPar:18,MozSaaBenDeb:17,YuanFenXu:18,XiaLuXu:18,VanChiPol:16,LinYajMur:18} for the optimal positioning and trajectory of the \ac{UAV}.
The optimal deployment of the \ac{UAV} has been presented to maximize the communication coverage according to system parameters in \cite{MozSaaBenDeb:15,MozSaaBen:16,CheDhi:17,AlzYan:18,WanZhaTia:18,CheAlzYan:19,KimLeeQue:19,AzaRosPol:19,AzaGerGar:19}.
The user scheduling and the \ac{UAV} trajectory have been jointly optimized to maximize the minimum average rate in \cite{WuZenZha:18} and the minimum secrecy rate in \cite{LeeEomPar:18}.
The \ac{UAV} trajectory is also optimized jointly with the device-\ac{UAV} association and the uplink power to minimize the total transmit power according to the number of update times in \cite{MozSaaBenDeb:17}.
The random 3D trajectory of the \ac{UAV} has been presented to maximize the link capacity between the \acp{UAV} in \cite{YuanFenXu:18}.
The work in \cite{XiaLuXu:18} proposed an anti-jamming relay strategy for the \ac{UAV}-aided \ac{VANET}.
The performance of the \ac{UAV} communication over the \ac{LTE} network has been analyzed by the measurement and simulation results in \cite{VanChiPol:16,LinYajMur:18}.
However, all of those prior works considered limited \ac{UAV} communication scenarios or environments.
Specifically, only the path loss is used for channels without fading in \cite{MozSaaBenDeb:15,MozSaaBen:16,WuZenZha:18,LeeEomPar:18,MozSaaBenDeb:17,XiaLuXu:18},
or the fact that the \ac{LoS} probability can be different according to the locations of the \ac{UAV} was not considered in \cite{CheDhi:17,YuanFenXu:18}.
In addition, the works in \cite{AlzYan:18,WanZhaTia:18,CheAlzYan:19,KimLeeQue:19,AzaRosPol:19,AzaGerGar:19} considered the different channel fadings depending on the \ac{LoS} probability. However, the works in \cite{AlzYan:18,WanZhaTia:18,CheAlzYan:19,KimLeeQue:19,AzaRosPol:19} used the path loss exponents and channel fading parameters, which are constant, not changed by the horizontal distance and the vertical distance of the communication link. The work in \cite{AzaGerGar:19} used the different path loss exponents according to the \ac{UAV} height, while the channel fading parameters are constant.

\begin{table}
	\caption{Notations used throughout the paper.} \label{table:notation}
	\begin{center}
		\rowcolors{2}
		{cyan!15!}{}
		\renewcommand{\arraystretch}{1.3}
		\begin{tabular}{ c  p{6cm} }
			\hline 
			{\bf Notation} & {\hspace{2cm}}{\bf Definition}
			\\
			\midrule
			\hline
			$i \in  \{\text{m}, \text{I}\}$ & Index for the main link ($i$ = m) and the interference link ($i$ = I) \\ \addlinespace
			$h_i$ & Channel fading gain of the link $i$\\ \addlinespace
			$\ell_i$ & Distance of the link $i$ \\ \addlinespace
			$\mathcal{D}=(\ell_\text{m},\ell_\text{I})$ & Link distance set \\ \addlinespace
			$d_i^{(\text{H})}$ 	& Horizontal distance of the link $i$ \\ \addlinespace
			$d_i^{(\text{V})}$ 	& Vertical distance of the link $i$ \\ \addlinespace
			$\alpha(\ell_i)$ & Path loss exponent of the link $i$ for given $\ell_i$\\ \addlinespace
			$K(\ell_i)$ & Rician factor for given $\ell_i$\\ \addlinespace
			$\PLoS$ & LoS probability for given $\ell_i$ \\ \addlinespace
			$P_i$ & Transmission power of the link $i$\\ \addlinespace
			$N_\text{o}$ & Noise power\\ \addlinespace
			$\gamma(\ell_\text{m},\ell_\text{I})$ & Signal-to-interference-plus-noise ratio (SINR) \\ \addlinespace
			$\hat{\gamma}(\ell_\text{m},\ell_\text{I})$ & Signal-to-interference ratio (SIR) \\ \addlinespace
			$\gamma_\text{t}$ & Target SINR/SIR \\ \addlinespace
			$e_i \in  \{\text{L}, \text{N}\}$ & Index for the LoS environment ($e_i$ = L) and the NLoS environment ($e_i$ = N) \\ \addlinespace
			$\Pout^{({e_\text{m},e_\text{I}})}(\mathcal{D})$	& Outage probability with the environment of the main link $e_\text{m}$ and that of the interference link $e_\text{I}$\\
			\hline 
		\end{tabular}
	\end{center}\vspace{-0.63cm}
\end{table}%
%
%
%
%
Therefore, in this paper,
we analyze the effect of an interfering node on the \ac{UAV} communications by considering both the \ac{LoS} and \ac{NLoS} links and channel fading.
We consider more realistic channel model for \ac{UAV} communications.
Specifically, the probability of forming the \ac{LoS} link is determined by the heights of the transmitter and the receiver and the horizontal and vertical distances of communication links.
Not only the pathloss exponent but also the fading channel factors (e.g., Rician factor) are modeled to be affected by the \ac{LoS} probability.
%
%
The main contribution of this paper can be summarized as follows:
\begin{itemize}
	\item we consider all possible scenarios of the main (i.e., from a transmitter to a receiver) and the interference (i.e., from an interfering node to a receiver) links on \ac{UAV} communications, of which channels can be \ac{G2A}, \ac{G2G}, \ac{A2G}, or \ac{A2A} channels;
	\item we provide the outage probability in the presence of an interfering node for all the scenarios in general environments by considering the \ac{LoS} probability and different channel fadings for \ac{LoS} and \ac{NLoS} links;
	\item we derive a \emph{closed-form} outage probability for the interference-limited environments, and using it, we also figure out whether the \ac{LoS} environments for both main and interference links can be better than the \ac{NLoS} environments in terms of the outage probability;
	\item we then analyze how the outage probability is affected by the heights of a transmitter and an interfering node and the link distances, and show the optimal \ac{UAV} heights that minimize the outage probability through numerical results; and
	\item we finally present the network outage probability by considering a network with multiple transmitting (also interfering) nodes and a UAV receiver in the air, 
	and show the effect of the transmitting node density on the optimal \ac{UAV} height.
\end{itemize}

The remainder of this paper is organized as follows.
In Section \ref{sec:models},
we present the network model and the channel model affected by horizontal and vertical distances of communication links.
We then derive a closed-form outage probability for the general environment and the interference-limited environment in Section \ref{sec:analytical}.
In Section \ref{sec:networkoutage},
we present the network outage probability considering multiple transmitting (also interfering) nodes.
In Section \ref{sec:numerical},
we evaluate the performance of \ac{UAV} communications according to the \ac{UAV} height, system parameters, and the channel environment. 
We then compare the optimal \ac{UAV} heights of the multiple interfering nodes case with that of the single nearest interfering node case.
Finally, the conclusion is presented in Section \ref{sec:conclusion}.

{\it Notation}: The notation used throughout the paper is reported in Table~\ref{table:notation}. 
%
%
%
%
\section{System Model}\label{sec:models}
In this section, we describe the network model and the channel model on \ac{UAV} communications.
\subsection{Terrestrial \& Aerial Network Models}
We consider a \ac{UAV} network, which has a \ac{UAV}, a ground device (e.g., ground control station or base station), and an interfering node.
In this network, 
there can be three types of communications: \ac{UAV} to \ac{UAV}, 
\ac{UAV} to ground device (or ground device to \ac{UAV}), and ground device to ground device.
The interfering node can be either on the ground or in the air, and we consider one interfering node.\footnote{Note that the result of this paper can be readily extended for the multiple interfering nodes case as presented in Section \ref{sec:networkoutage}.
However, the analysis results will be complicated and give fewer insights.
In addition, the communication performance is generally determined by one critical interfering node, especially in low outage probability region \cite{MorLoy:09}.
Therefore, we focus on the one interfering node case in this work, 
but the performance for the multiple interfering nodes case is also presented in simulation results of Section \ref{sec:numerical}.}

%
\begin{figure}[t!]
	\begin{center}   
		{ 
			\includegraphics[width=1.00\columnwidth]{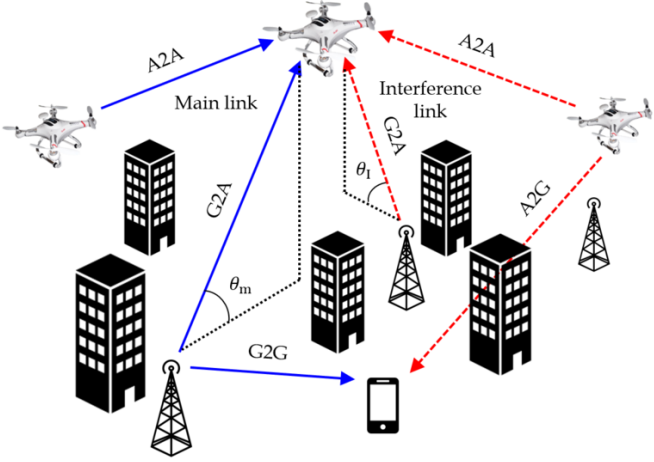}
			\vspace{-10mm}
		}
	\end{center}
	\caption{
		System model when UAVs are the communication devices. There are four types of channels: ground-to-ground (G2G), ground-to-air (G2A), air-to-ground (A2G), and air-to-air (A2A) channels. The blue lines represent the main links and the red dotted lines represent the interference links.
	}
	\label{fig:system}
\end{figure}
%
\par
When a transmitter (Tx), located at $(x_\text{m}, y_\text{m}, z_\text{m})$, communicates to a receiver (Rx), located at $(0,0, z_\text{o})$, 
in the presence of an interfering node at $(x_\text{I}, y_\text{I}, z_\text{I})$, 
\ac{SINR} is given by
\begin{align} \label{eq:SINR}
	\gamma(\ell_\text{m},\ell_\text{I})&=
	\frac{\Hm \Lma \Pm}  {\Hi \Lia \PI + N_\text{o}} 
	= \frac{\Hm \beta_\text{m}(\ell_\text{m})}  {\Hi \beta_\text{I}(\ell_\text{I}) + N_\text{o}}
\end{align}
where $\beta_\text{m}(\ell_\text{m})$ and $\beta_\text{I}(\ell_\text{I})$ are respectively given by 
\begin{align}	
	\Beta{m}=\Lma \Pm, \quad
	\Beta{I}=\Lia \PI.
\end{align}
Here, $\Hm$ and $\Hi$ are the fading gains of the main link (i.e., the channel between Tx and Rx) and the interference link (i.e., the channel between interfering node and Rx), respectively; $\Lm\hspace{-0.7mm}=\hspace{-0.7mm}\sqrt{x_\text{m}^2+y_\text{m}^2+(z_\text{m}-z_\text{o})^2}$ and
$\Li\hspace{-0.7mm}=\hspace{-0.7mm}\sqrt{x_\text{I}^2+y_\text{I}^2+(z_\text{I}-z_\text{o})^2}$ are the distances of the main link and the interference link, respectively;
$\Pm$ and $\PI$ are the transmission power of the Tx and the interfering node, respectively; $\PL{\text{m}}(\ell_\text{m})$ and $\PL{\text{I}}(\ell_\text{I})$ are the path loss exponents of the main link and the interference link, respectively; and $N_\text{o}$ is the noise power.
Here,
we define that $d_i^{(\text{H})} = \sqrt{x_i^2 + y_i^2}$ is the horizontal distance
and $d_i^{(\text{V})} = \sqrt{(z_i - z_\text{o})^2}$ is the vertical distance of the main link ($i = \text{m}$) or the interference link ($i = \text{I}$).
\subsection{Channel Model} \label{subsec:channel}
As shown in Fig. \ref{fig:system}, 
there are three types of the channels in the UAV networks: 
the \emph{\ac{A2G} channel} (from \ac{UAV} to a ground device), the \emph{\ac{A2A} channel} (from \ac{UAV} to \ac{UAV}), and the \emph{\ac{G2G} channel} (from a ground device to a ground device).
The \ac{G2G} channel is the same channel of a terrestrial network, which is generally modeled as the \ac{NLoS} environment with Rayleigh fading in urban areas. 
The \ac{G2A} channel and the \ac{A2G} channel have the same characteristics, so we describe characteristics of the \ac{A2G} and \ac{A2A} channels in this subsection.

The \ac{A2G} and \ac{A2A} channels can have the \ac{LoS} or \ac{NLoS} environment depending on the height of the \ac{UAV} and its surrounding environment such as buildings.
In the following, 
we describe the channel components affected by the horizontal distance $d_i^{(\text{H})}$ and the vertical distance $d_i^{(\text{V})}$, and then provide the models for \ac{A2G} and \ac{A2A} channels.

\subsubsection{Channel components} \label{subsec:component}
\begin{figure}[t!]
	\begin{center}   
		{ 
			\psfrag{AAAAAAAAAAAAAAAAAAAAAAAAA11}[Bl][Bl][0.59]{Rician factor in \cite{AzaRoChePol:18}}
			\psfrag{A2}[Bl][Bl][0.59]{Proposed Rician factor $(0m-200m)$}
			\psfrag{A3}[Bl][Bl][0.59]{Proposed Rician factor $(20m-220m)$}
			\psfrag{X}[Bl][Bl][0.59]{$\text{Horizontal distance of main link},\: d_\text{m}^{(\text{H})} \:[m]$}
			\psfrag{Y}[Bl][Bl][0.59]{Rician factor, $K(\ell_\text{m})$}
			\includegraphics[width=1.00\columnwidth]{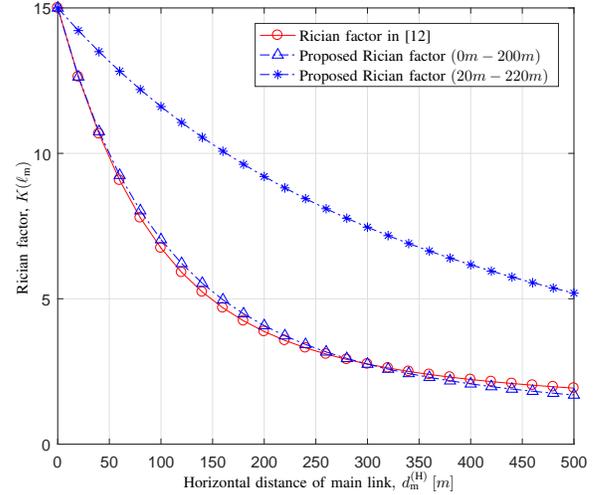}
			\vspace{-6mm}
		}
	\end{center}
	\caption{
		Rician factor $K(\ell_\text{m})$ as a function of $\Dist{m}{H}$ with $d_\text{m}^{(\text{V})}=200m$.}
	\label{fig:Rician}
\end{figure}

The horizontal distance $d_i^{(\text{H})}$ and the vertical distance $d_i^{(\text{V})}$ of the communicatin link affect the probability of forming \ac{LoS} link, the path loss exponent, and the Rician factor as described below.
%
%
%
\bi
\item 
The \emph{\ac{LoS} probability} is given by \cite{BaiVazHea:14,YanZhoZha:18}
\begin{align}
&\PLoS =	\label{eq:LoS} \\
&\left\{
\begin{aligned}
&
\left\{ 1 - \exp \left( - \frac{z_i^2} {2\zeta^2} \right) \right\}^{\ell_i \sqrt{\nu \mu} } \quad\quad\quad\quad\quad\,\,\,\,\,\text{for} \,\,   z_i = z_\text{o} \nonumber \\
&
\left\{ 1 \hspace{-0.5mm} - \hspace{-0.5mm} \frac{\sqrt{2\pi} \zeta} {d_i^{(\text{V})}} 
\Bigg| Q\hspace{-1mm}\left(\frac{z_i} {\zeta}\right) \hspace{-1mm} - \hspace{-0.5mm} Q\hspace{-1mm}\left(\frac{z_\text{o}} {\zeta}\right) \hspace{-1mm} \Bigg| \right\}^{d_i^{(\text{H})} \hspace{-0.5mm}\sqrt{\nu \mu} } \,\,\,\text{for} \,\,  z_i \ne z_\text{o} 
\end{aligned}
\right.
\end{align}
where $Q(x)=\int_{x}^{\infty} \frac{1} {\sqrt{2\pi}} \exp \left(- \frac{t^2} {2}\right) \,dt$ is the Q-function and $\zeta$, $\nu$, and $\mu$ are environment parameters, which are determined by the building density and heights of the Tx and the Rx.
Furthermore, the \ac{NLoS} probability is $\PNLoS=1-\PLoS$.
\item 
The \emph{path loss exponent} is determined by $\ell_i$ as \cite{AzaRoChePol:18}
\begin{align} \label{eq:pathloss}
\alpha(\ell_i)=
a_1 \PLoS + b_1
\end{align}
where 
$a_1=
\alpha_\text{L}  -  \alpha_\text{N}$ and  
$b_1=
\alpha_\text{N}$.
Here, $\alpha_\text{L}$ and $\alpha_\text{N}$ are the path loss exponenets when the \ac{LoS} probabilities are one and zero, respectively.
%
%
%
%
%
\item 
The \emph{Rician factor} is proposed to be determined by $\ell_i$ as
\begin{align}	\label{eq:k-factor}
K(\ell_i)
= a_2 \exp\left\{b_2 \PLoS^2\right\}
\end{align}
where 
$a_2=K_\text{N}$ and
$b_2= 
\ln\left(\frac{K_\text{L}}  {K_\text{N}}\right)$.
Here, $K_\text{L}$ and $K_\text{N}$ are denoted as the Rician factors when the \ac{LoS} probabilities are one and zero, respectively.
Note that the Rician factor, defined by the elevation angle $\theta_i$ 
as $K(\theta_i)= a_2 \exp(b_2 \theta_i)$ \cite{AzaRoChePol:18}, was used in prior works. 
However, this model has a problem when applied to the \ac{A2A} channel.
For example, due to smaller elevation angle of \ac{A2A} channel, 
the Rician factor of the \ac{A2A} channel becomes 
smaller than that of the \ac{A2G} channel.
This means the average channel fading gain of the \ac{A2A} channel is smaller than 
that of the \ac{A2G} channel, which is not true in reality. 
On the other hand, the proposed Rician factor model in \eqref{eq:k-factor} 
is changed 
according to the respective heights of the receiver and the transmitter
as shown in Fig.~\ref{fig:Rician}.
%
%

\hspace{10pt}Figure~\ref{fig:Rician} presents the Rician factors $K(\ell_\text{m})$ as a function of the horizontal distance of main link $\Dist{m}{H}$ for different values of $z_\text{o}$ and $z_\text{m}$, where the Rx is located at $(0,0,z_\text{o})$ and the Tx moves from $(0,0,z_\text{m})$ to $(x_\text{m},y_\text{m},z_\text{m})$.
	From this figure, we can see that the Rician factor decreases with $\Dist{m}{H}$ because the \ac{LoS} probability decreases with $\Dist{m}{H}$.
	We can also see that the Rician factor of the \ac{A2A} channel (i.e., $20m-220m$) is greater than that of the \ac{A2G} channel (i.e., $0m-200m$) even though the elevation angles of both \ac{A2A} and \ac{A2G} channels are the same.
	In addition, the proposed Rician factor has similar trend to the Rician factor in \cite{AzaRoChePol:18} with the same simulation environment.
%
%
\ei
Note that from \eqref{eq:LoS}-\eqref{eq:k-factor}, we can see that $\PLoS$ and $K(\ell_i)$ are increasing functions of $d_i^{(\text{V})}$ and $\alpha(\ell_i)$ is a decreasing function of $d_i^{(\text{V})}$, so the received power increases as $d_i^{(\text{V})}$ increases.
\subsubsection{Air-to-Ground (A2G) \& Air-to-Air (A2A) channels}
When the main link and the interference link are \ac{A2G} or \ac{A2A} channel, $\Hm$ and $\Hi$ can be in either the \ac{LoS} or \ac{NLoS} environment. 
We consider that the channel fading is Rician fading for the \ac{LoS} environment and Rayleigh fading for the \ac{NLoS} environment.
Therefore, 
the distribution of the channel fading, $h_i$, $i \in  \{\text{m}, \text{I}\}$, is given by
\begin{align}
	f_{h_i}(h)=	
	\left\{
		\begin{aligned}
		&
		f_\text{L}(h) \quad\quad \text{for \ac{LoS} case} \label{eq:channel fading}\\
		&
		f_\text{N}(h) \quad\quad \text{for \ac{NLoS} case}
		\end{aligned}
	\right.
\end{align}
where $f_\text{L}(h)$ and $f_\text{N}(h)$ are noncentral Chi-squared and exponential distribution, respectively, and given by
\begin{align}	
	f_\text{L}(h)
		&=\frac{1 + K(\ell_i)}  {\HL}  
		\exp\left(-K(\ell_i)-\frac{1+K(\ell_i)}  {\HL} h \right)\nonumber\\
		&\quad\times I_0 \left(2\sqrt{\frac{K(\ell_i) (1 + K(\ell_i)) }  {\HL} h} \right)\nonumber\\
		&=\frac{1}  {2}  
		\exp\left(-K(\ell_i) - \frac{h}  {2}\right) 
		I_0 \left(\sqrt{2 K(\ell_i) h} \right) \label{eq:channel fading1}\\
	f_\text{N}(h)
		&=\frac{1}  {\HN}
		\exp\left(-\frac{h}  {\HN}\right)
		=\exp\left(-h\right). \label{eq:channel fading2}
\end{align}
Here, $I_0(\cdot)$ is the modified Bessel function of the first kind with order zero,
and $\HL=2+2K(\ell_i)$ and $\HN=1$ are the means of \ac{LoS} and \ac{NLoS} channel fading gain, respectively.

\setcounter{eqnback}{\value{equation}}
\setcounter{equation}{10}
\begin{figure*}[t!]
	\begin{align}
	%
	%
	\Pout^{(\text{L,L})}(\mathcal{D}) &=
	1 - \frac{1}  {2} \int_{0}^{\infty}
	Q\left(\sqrt{2 \Km(\ell_\text{m})} , 
	\sqrt{\frac{\gamma_\text{t} (\Beta{I} g + N_\text{o})} {\Beta{m}}}\right)
	\exp\left(-\Ki(\ell_\text{I}) - \frac{g}  {2}\right)
	I_0\left(\sqrt{2 \Ki(\ell_\text{I})  g}\right) \, dg  \label{eq:Po1} \\
%
%
	\Pout^{(\text{L,N})}(\mathcal{D}) &=
	1 - Q\left(\sqrt{2 \Km(\ell_\text{m})} , 
	\sqrt{\frac{\gamma_\text{t} N_\text{o}}  {\Beta{m}} }\right)  
	+ \frac{\gamma_\text{t} \Beta{I}}  {2 \Beta{m} + \gamma_\text{t} \Beta{I}} 
	\exp\left(\frac{N_\text{o}}  {\Beta{I}} - \frac{2 \Km(\ell_\text{m}) \Beta{m}}  {2 \Beta{m} + \gamma_\text{t} \Beta{I}}\right)  \nonumber \\
	&\quad\times
	Q\left(\sqrt{\frac{2 \gamma_\text{t} \Km(\ell_\text{m}) \Beta{I}}  
		{2 \Beta{m} + \gamma_\text{t} \Beta{I}}} , 
	\sqrt{\frac{N_\text{o}(2 \Beta{m} + \gamma_\text{t} \Beta{I})}  
		{\Beta{m} \Beta{I}}}\right)
	\label{eq:Po2}
	\end{align}
	
	\setcounter{eqncnt}{12}
	\centering \rule[0pt]{18cm}{0.3pt}
\end{figure*}
\setcounter{equation}{\value{eqnback}}
%
%
%
\section{Outage Probability Analysis}\label{sec:analytical}
In this section, 
we analyze the outage probability of \ac{UAV} communications by considering various environments of main and interference links. 
The outage probability is provided for two cases:
the general environment in Section \ref{sec:SINR} and
the interference-limited environment in Section \ref{sec:SIR}. 
\subsection{General Environments}\label{sec:SINR}
%
%
%
For given the link distance set $\mathcal{D} =  (\ell_\text{m},\ell_\text{I})$ of main and interference links,
the outage probability is defined as
\begin{align} \label{eq:out}
	\Pout(\mathcal{D})=
		\mathbb{P}[\gamma(\ell_\text{m},\ell_\text{I}) < \gamma_\text{t}]
\end{align}
where $\gamma_\text{t}$ is the target \ac{SINR} or \ac{SIR}, 
which can be defined by $\gamma_\text{t}=2\!^{\frac{R_\text{t}}  {W}} - 1$ for the target rate $R_\text{t}$ and the bandwidth $W$\cite{LimYaoMan:09,IkkAhm:09,GraRyz:B07}.
Using \eqref{eq:out}, the outage probability can be derived from the distribution of the channel fading as follows.
\begin{theorem}\label{trm:OPSINR}
For given $\mathcal{D} =  (\ell_\text{m},\ell_\text{I})$, 
the outage probability $\Pout(\mathcal{D})$ can be presented as
\begin{align} \label{eq:outage}
	\Pout(\mathcal{D})
		&=
		\sum_{e_\text{m}, e_\text{I} \in \{\text{L},\text{N}\}}^{}
		p_{e_\text{m}}(\ell_\text{m}) p_{e_\text{I}}(\ell_\text{I}) \Pout^{(e_\text{m},e_\text{I})}(\mathcal{D}) 
		\nonumber\\
		&=
		\PLoSm \PLoSi \Pout^{(\text{L,L})}(\mathcal{D}) 
		\nonumber\\
		& \quad +
		\PLoSm \PNLoSi 
		\Pout^{(\text{L,N})}(\mathcal{D}) \nonumber\\
		& \quad +
		\PNLoSm \PLoSi 
		\Pout^{(\text{N,L})}(\mathcal{D}) \nonumber\\
		& \quad + 
		\PNLoSm \PNLoSi
		\Pout^{(\text{N,N})}(\mathcal{D})
\end{align}
where $\Pout^{({e_\text{m},e_\text{I}})}(\mathcal{D})$ is the outage probability 
with the environment of the main link $e_\text{m}$ and that of the interference link $e_\text{I}$.
The environment $e_i$ can be either LoS (i.e., $e_i = \text{L}$) or NLoS (i.e., $e_i = \text{N}$), and 
$\Pout^{({e_\text{m},e_\text{I}})}(\mathcal{D})$ for four cases of $({e_\text{m},e_\text{I}})$ are given as follows:
\be
\item \emph{Case} 1 ($e_\text{m}=\text{L}$ and $e_\text{I}=\text{L}$):
$\Pout\!^{(\text{L,L})}(\mathcal{D})$ is given by \eqref{eq:Po1}.
\item \emph{Case} 2 ($e_\text{m}=\text{L}$ and $e_\text{I}=\text{N}$):
$\Pout\!^{(\text{L,N})}(\mathcal{D})$ is given by \eqref{eq:Po2}.
\item \emph{Case} 3 ($e_\text{m}=\text{N}$ and $e_\text{I}=\text{L}$):
$\Pout\!^{(\text{N,L})}(\mathcal{D})$ is given by
%
%
\setcounter{equation}{12}
\begin{align}
	\Pout\!^{(\text{N,L})}\hspace{-0.3mm}(\mathcal{D}) \hspace{-0.5mm} & = \hspace{-0.5mm}
	1  -  \frac{\Beta{m}}  {2 \gamma_\text{t} \Beta{I} + \Beta{m}}  \nonumber \\
	&\hspace{-0.5mm}\quad\times \hspace{-0.5mm}
	\exp \hspace{-0.5mm}
		\left( \hspace{-1mm} 
			- \frac{\gamma_\text{t} N_\text{o}} 
				{\beta_\text{m}(\hspace{-0.3mm} \ell_\text{m} \hspace{-0.3mm})} 
		\hspace{-0.5mm} - \hspace{-0.5mm}
			\frac{2 \gamma_\text{t} \Ki(\hspace{-0.3mm} \ell_\text{I} \hspace{-0.3mm}) \beta_\text{I}(\hspace{-0.3mm} \ell_\text{I} \hspace{-0.3mm})}  
				{2 \gamma_\text{t} \beta_\text{I}
					(\hspace{-0.3mm} \ell_\text{I} \hspace{-0.3mm}) 
					\hspace{-0.5mm} + \hspace{-0.5mm}  
					\beta_\text{m}(\hspace{-0.3mm} \ell_\text{m} \hspace{-0.3mm})}
		\hspace{-0.8mm}\right) 
		\hspace{-1.2mm}.
	\hspace{-1.6mm} \label{eq:Po3}
\end{align}
\item \emph{Case} 4 ($e_\text{m}=\text{N}$ and $e_\text{I}=\text{N}$):
$\Pout\!^{(\text{N,N})}(\mathcal{D})$ is given by
%
%
\begin{align}
	\Pout\!^{(\text{N,N})}\hspace{-0.3mm}(\mathcal{D}) \hspace{-0.5mm} = \hspace{-0.5mm}
	1 \hspace{-0.5mm} - \hspace{-0.5mm} \frac{\Beta{m}}  
	{\Beta{m} \hspace{-0.6mm} + \hspace{-0.6mm} \gamma_\text{t} \Beta{I}}
	\hspace{-0.5mm}
	\exp\hspace{-0.5mm}\left(\hspace{-1mm} - \frac{\gamma_\text{t} N_\text{o}}  {\Beta{m}}\hspace{-0.5mm}\right)\hspace{-1mm}.  
	\hspace{-1.5mm} \label{eq:Po4}
\end{align}
\ee
\end{theorem}
\begin{IEEEproof}
See Appendix~\ref{app:trm}.
\end{IEEEproof}

From Theorem \ref{trm:OPSINR}, we can also obtain the outage probability for different scenarios of \ac{UAV} communications by changing the values of $(z_\text{m},z_\text{I},z_\text{o})$.
Specifically, when the \ac{LoS} probabilities of main and interference links increase to one according to the values of $(z_\text{m},z_\text{I},z_\text{o})$,
it is only necessary to consider the outage probability $\Pout^{(\text{L,L})}(\mathcal{D})$ in \eqref{eq:Po1}.

%
\setcounter{eqnback}{\value{equation}}
\setcounter{equation}{16}
\begin{figure*}[t!]
	%
	%
	\begin{align}
	\Pouth\!^{(\text{L,L})}(\mathcal{D}) 
	&=  1  -  Q\left(\sqrt{\frac{2 \Km(\ell_\text{m}) \Beta{m}}  
		{\Beta{m} + \gamma_\text{t} \Beta{I}}}  , 
	\sqrt{\frac{2 \gamma_\text{t} \Ki(\ell_\text{I}) \Beta{I}}  
		{\Beta{m} + \gamma_\text{t} \Beta{I}}}\right)
	+ \frac{\gamma_\text{t} \Beta{I}}  {\Beta{m} + \gamma_\text{t} \Beta{I}} \nonumber \\
	&\quad\times\exp\left(-\frac{\Km(\ell_\text{m}) \Beta{m} 
		+ \gamma_\text{t} \Ki(\ell_\text{I}) \Beta{I}}  
	{\Beta{m} + \gamma_\text{t} \Beta{I}} \right) 
	I_0\left(\frac{2 \Beta{m}}  {\Beta{m} + \gamma_\text{t} \Beta{I}}	
	\sqrt{\frac{\gamma_\text{t} \Km(\ell_\text{m}) \Ki(\ell_\text{I}) \Beta{I}}  {\Beta{m}}} \right)  \label{eq:Po1h}
	\end{align}
	\setcounter{eqncnt}{17}
	\centering \rule[0pt]{18cm}{0.3pt}
\end{figure*}
\setcounter{equation}{\value{eqnback}}
%
%
\subsection{Interference-limited Environments}\label{sec:SIR}
In this subsection, we provide the outage probability when it is dominantly determined by the received power of the interfering signal, i.e., the interference-limited environment.
We provide the outage probability in closed-forms, and they can also provide more insights on the effects of environments parameters on the outage probability.

In the interference-limited environment, the outage probability is defined as
\begin{align}\label{eq:outage_h}
	\Pouth(\mathcal{D})=
		\mathbb{P}[\hat{\gamma}(\ell_\text{m},\ell_\text{I}) < \gamma_\text{t}]
\end{align}
where $\hat{\gamma}(\ell_\text{m},\ell_\text{I})$ is the \ac{SIR}, given by
\begin{align} \label{eq:SIR}
	\hat{\gamma}(\ell_\text{m},\ell_\text{I})
		&=
		\frac{\Hm \Lma \Pm}  {\Hi \Lia \PI}
		=\frac{\Hm \Beta{m}}  {\Hi \Beta{I}}.
\end{align}
The outage probability can be derived by a similar approach in Theorem \ref{trm:OPSINR}, and provided in the following lemma.
\begin{lemma}\label{lem:OPSIR}
For given $\mathcal{D} =  (\ell_\text{m},\ell_\text{I})$, 
the outage probability $\Pouth(\mathcal{D})$ can be presented as \eqref{eq:outage} by substituting from $\Pout^{({e_\text{m},e_\text{I}})}(\mathcal{D})$ to $\Pouth^{({e_\text{m},e_\text{I}})}(\mathcal{D})$,
where $\Pouth^{({e_\text{m},e_\text{I}})}(\mathcal{D})$ are given as follows:
\be
\item \emph{Case} 1 ($e_\text{m}=\text{L}$ and $e_\text{I}=\text{L}$):
$\Pouth\!^{(\text{L,L})}(\mathcal{D})$ is given by \eqref{eq:Po1h}.
\setcounter{equation}{17}
\item \emph{Case} 2 ($e_\text{m}=\text{L}$ and $e_\text{I}=\text{N}$):
$\Pouth\!^{(\text{L,N})}(\mathcal{D})$ is given by
\begin{align}
	\Pouth\!^{(\text{L,N})}(\mathcal{D}) 
	&=
	\frac{\gamma_\text{t} \Beta{I}}  
	{2 \Beta{m}  +  \gamma_\text{t} \Beta{I}}
	\nonumber \\
	& \quad \times \exp\left( - \frac{2 \Km(\ell_\text{m}) \Beta{m}}  
	{2 \Beta{m}  +  \gamma_\text{t} \Beta{I}}\right).	\label{eq:Po2h}
\end{align}
\item \emph{Case} 3 ($e_\text{m}=\text{N}$ and $e_\text{I}=\text{L}$):
$\Pouth\!^{(\text{N,L})}(\mathcal{D})$ is given by
\begin{align}
	\Pouth\!^{(\text{N,L})}(\mathcal{D})
	 & =
	1 - \frac{\Beta{m}}  
	{2 \gamma_\text{t} \Beta{I}  +  \Beta{m}} \nonumber \\
	&\quad\times 
	\exp \left( - \frac{2 \gamma_\text{t} \Ki(\ell_\text{I}) \Beta{I}}  
	{2 \gamma_\text{t} \Beta{I} + \Beta{m}}\right) .	\label{eq:Po3h}
\end{align}
\item \emph{Case} 4 ($e_\text{m}=\text{N}$ and $e_\text{I}=\text{N}$):
$\Pouth\!^{(\text{N,N})}(\mathcal{D})$ is given by
\begin{align}
	\Pouth\!^{(\text{N,N})}(\mathcal{D}) = 
	\frac{\gamma_\text{t} \Beta{I}}  
	{\Beta{m} + \gamma_\text{t} \Beta{I}}.  \label{eq:Po4h}
\end{align}
\ee
\end{lemma}
\begin{IEEEproof}
See Appendix~\ref{app:lem}.
\end{IEEEproof}

From Lemma \ref{lem:OPSIR}, 
we can also obtain the outage probability for different scenarios of \ac{UAV} communications by changing the values of $(z_\text{m},z_\text{I},z_\text{o})$.

From Theorem \ref{trm:OPSINR} and Lemma \ref{lem:OPSIR}, 
we can readily know that $\Pout\!^{(\text{L,N})}(\mathcal{D})$ (Case 2) cannot be higher than $\Pout\!^{(\text{N,L})}(\mathcal{D})$ (Case 3) as Case 2 has stronger main link and weaker interference link than Case 3. 
However,
it is not clear whether the outage probability with LoS environments for both main and interference links (Case 1) can be lower or higher than that with NLoS environments for both main and interference links (Case 4).
Hence, we compare $\Pout\!^{(\text{L,L})}(\mathcal{D})$ and $\Pout\!^{(\text{N,N})}(\mathcal{D})$, and obtain the following results in Corollary 1.
%
%
%
\begin{corollary} \label{pro:outage compare}
	According to the ratio of the average received signal power of the main and interference links, i.e., $\frac{\beta_\text{m}(\ell_\text{m})}  {\beta_\text{I}(\ell_\text{I})}$, the relation between $\Pouth\!^{(\text{L,L})}(\mathcal{D})$ and $\Pouth\!^{(\text{N,N})}(\mathcal{D})$ is changed as
	\begin{align}\label{eq:Compare}	
	\left\{
	\begin{aligned}
	&
	\Pouth\!^{(\text{L,L})}(\mathcal{D}) > 
	\Pouth\!^{(\text{N,N})}(\mathcal{D}), 
	\,  \, \, \, \, \text{for} \, \, \, 0 < \frac{\beta_\text{m}(\ell_\text{m})}  {\beta_\text{I}(\ell_\text{I})}  <  v'
	\\
	&
	\Pouth\!^{(\text{L,L})}(\mathcal{D}) < 
	\Pouth\!^{(\text{N,N})}(\mathcal{D}), 
	\, 	\, \, \, \, \text{for} \, \, \,  v' < \frac{\beta_\text{m}(\ell_\text{m})}  {\beta_\text{I}(\ell_\text{I})} < \infty
	\\
	&
	\Pouth\!^{(\text{L,L})}(\mathcal{D})  =   
	\Pouth\!^{(\text{N,N})}(\mathcal{D}),
	\,  \, \, \, \, \text{for} \, \, \, \frac{\beta_\text{m}(\ell_\text{m})}  {\beta_\text{I}(\ell_\text{I})} = 0  , \infty, \, \text{or} \ v' \hspace{-1mm}
	\end{aligned}
	\right.
	\end{align}
	where $v'$ ($0< v' <\infty$) is the value of $\frac{\beta_\text{m}(\ell_\text{m})}  {\beta_\text{I}(\ell_\text{I})}$ 
	that makes $\Pouth\!^{(\text{L,L})}(\mathcal{D})  =   
	\Pouth\!^{(\text{N,N})}(\mathcal{D})$.
\end{corollary}
\begin{IEEEproof}
For convenience, 
we introduce $v=\frac{\beta_\text{m}(\ell_\text{m})}  {\beta_\text{I}(\ell_\text{I})}$,
and define $A(v)$ and $B(v)$ as 
\begin{align}
	A(v)=\sqrt{\frac{2 \Km(\ell_\text{m}) v}  {v + \gamma_\text{t} }}  , \quad
	B(v)=\sqrt{\frac{2 \gamma_\text{t} \Ki(\ell_\text{I}) }  {v + \gamma_\text{t} }}.
\label{eq:new}
\end{align}
%
%
%
By using \eqref{eq:new}, 
$\Pouth\!^{(\text{L,L})}(\mathcal{D}) $ in \eqref{eq:Po1h} and $\Pouth\!^{(\text{N,N})}(\mathcal{D})$ in \eqref{eq:Po4h} can rewrite as functions of $v$ as
\begin{align}
	\Pouth\!^{(\text{L,L})}(v) 
		& =  1  - Q\left(A(v), B(v)\right)
		+  \frac{\gamma_\text{t} }  {v + \gamma_\text{t} } \nonumber \\
		&\quad\times
		\exp\left( -\frac{A(v)^2  +  B(v)^2} {2}  \right) 
		I_0\left( A(v)B(v)  \right)   \nonumber \\
	\Pouth\!^{(\text{N,N})}(v) &=
		\frac{\gamma_\text{t} }  {v + \gamma_\text{t} }.  \label{eq:Poh_r}
\end{align}
From \eqref{eq:Poh_r}, 
we obtain the first derivatives of $\Pouth\!^{(\text{L,L})}(v)$ and $\Pouth\!^{(\text{N,N})}(v)$ according to $v$, respectively, as
\begin{align}\label{eq:dif2}
&\frac{\partial \Pouth\!^{(\text{L,L})}(v)} {\partial v} 
\hspace{-0.5mm} = \hspace{-0.5mm}
	\left(\Pouth\!^{(\text{N,N})}(v) \hspace{-0.5mm} - \hspace{-0.5mm} 1 \right)
\hspace{-0.5mm}	\exp \hspace{-0.5mm} \left(\hspace{-0.5mm} - \frac{A( v )^2  +  B( v )^2}  {2}  \right) \hspace{-0.5mm}
	B(v)
\nonumber \\
&\times 
	\Bigg\{ I_1 \left( A(v) B(v) \right) 
	\frac{\partial A( v )} {\partial v} 
	- I_0 \left(A(v) B(v)\right)
	\frac{\partial B(v)} {\partial v}  \Bigg\}
\nonumber \\
&	+  \Pouth\!^{(\text{N,N})}(v)
	\exp\left(-\frac{A(v)^2  +  B(v)^2}  {2}  \right) 
	A(v)
\nonumber \\
&\times
	\left\{I_1\left(A(v)B(v)\right)
	\frac{\partial B(v)}{\partial v}
	-  I_0\left(A(v)B(v)\right)
	\frac{\partial A(v)}{\partial v} \right\} \nonumber \\
& + \hspace{-0.5mm} 
\frac{\partial \Pouth\!^{(\text{N,N})}(v)} {\partial v} \hspace{-0.5mm}
	\exp\hspace{-0.8mm}\left(\hspace{-1mm}-\frac{A(v)^2 \hspace{-0.5mm} + 
	\hspace{-0.5mm} B(v)^2}  {2} \right) \hspace{-1mm}
	I_0\hspace{-0.5mm}\left(A(v)B(v)\right) \hspace{-0.5mm}
	< \hspace{-0.5mm} 0 \hspace{-1mm}
\end{align}
\vspace{-5mm}
\begin{align}
	\frac{\partial \Pouth\!^{(\text{N,N})}(v)} {\partial v} = 
	- \frac{\gamma_\text{t}} {\left(v + \gamma_\text{t}\right)^2} 
	< 0 .
\label{eq:dif1}
\end{align}
In \eqref{eq:dif2} and \eqref{eq:dif1}, 
the inequalities are obtained since
$\exp(v) \ge 1$, $I_0(v) \ge 1$, $A(v) \ge 0$, $B(v) \ge 0$, $I_1(v) \ge 0$, $\frac{\partial A(v)}{\partial v} \ge 0$, $\frac{\partial B(v)}{\partial v} \le 0$, and 
$0 \le \Pouth\!^{(\text{N,N})}(v) \le 1$.
Hence, we can see that $\Pouth\!^{(\text{L,L})}(v)$ and $\Pouth\!^{(\text{N,N})}(v)$ are
monotonically decreasing functions of $v$.
If $v = 0$, from \eqref{eq:dif2} and \eqref{eq:dif1}, we have
\begin{align}	
	\frac{\partial \Pouth\!^{(\text{N,N})}(0)} {\partial v} 
	<
	\frac{\partial \Pouth\!^{(\text{L,L})}(0)} {\partial v}
\end{align}
since 
$
\frac{\partial \Pouth\!^{(\text{N,N})}(0)} {\partial v} = 
- \frac{1} {\gamma_\text{t}},
$
$
\frac{\partial \Pouth\!^{(\text{L,L})}(0)} {\partial v} =  
\frac{\partial \Pouth\!^{(\text{N,N})}(0)} {\partial v} \exp\left(-\frac{B(0)^2}  {2} \right)
$,
and $\Pouth\!^{(\text{N,N})}(0) = \Pouth\!^{(\text{L,L})}(0) = 1$.
Hence, for small $\epsilon$, we have
\begin{align}\label{eq:relation1}		
	\Pouth\!^{(\text{N,N})}(\epsilon) < \Pouth\!^{(\text{L,L})}(\epsilon). 
\end{align}
If $v$ approaches $\infty$, $B(v) \rightarrow 0 $,
$\lim_{v \rightarrow \infty} \Pouth\!^{(\text{L,L})}(v) 
= 
\lim_{v \rightarrow \infty} \Pouth\!^{(\text{N,N})}(v)
= 0 
$, 
and from \eqref{eq:dif2} and \eqref{eq:dif1}, 
we have
\begin{align} \label{eq:dif3}
&\frac{\partial \Pouth\!^{(\text{N,N})}(v)} {\partial v}\rightarrow
- \frac{\gamma_\text{t}} {\left(v + \gamma_\text{t}\right)^2}, \nonumber \\
&\frac{\partial \Pouth\!^{(\text{L,L})}(v)} {\partial v} \rightarrow   
\frac{\partial \Pouth\!^{(\text{N,N})}(v)} {\partial v} \exp\left(-\frac{A(v)^2}  {2} \right).	%
\end{align}
From \eqref{eq:dif3}, we can see that for large $v_\text{o} \gg1$,
$\frac{\partial \Pouth\!^{(\text{L,L})}(v_\text{o})} {\partial v}
>
\frac{\partial \Pouth\!^{(\text{N,N})}(v_\text{o})} {\partial v}$, and we have
\begin{align} \label{eq:relation2}	
\Pouth\!^{(\text{L,L})}(v_\text{o}) < \Pouth\!^{(\text{N,N})}(v_\text{o})
\end{align}
Therefore, 
from \eqref{eq:relation1}, \eqref{eq:relation2}, and the fact that 
$\Pouth\!^{(\text{L,L})}(v)$ and $\Pouth\!^{(\text{N,N})}(v)$ are both monotonically decreasing functions, 
we can know that there exists unique point $v'$ in $0<v'<\infty$ that makes 
$\Pouth\!^{(\text{L,L})}(v')=\Pouth\!^{(\text{N,N})}(v')$. Therefore, we obtain \eqref{eq:Compare}.
\end{IEEEproof}

From Corollary \ref{pro:outage compare}, we can see that 
	when the main and interference links are in the same environment,
	the \ac{NLoS} environment can be preferred if the average received power of the interference is much larger than that of the transmitting signal (i.e., small $ \frac{\beta_\text{m}(\ell_\text{m})}  {\beta_\text{I}(\ell_\text{I})} $). However, for the opposite case (i.e., large $ \frac{\beta_\text{m}(\ell_\text{m})}  {\beta_\text{I}(\ell_\text{I})} $), the \ac{LoS} environment can be better 
	in terms of the outage probability.
	{This result will also be verified in numerical results of Section \ref{sec:environment}.}
\section{Network Outage Probability} \label{sec:networkoutage}
In this section,
we consider the interference-limited environment and the \ac{UAV} network where a receiving \ac{UAV} is in the air and 
multiple transmitting nodes are randomly distributed in \ac{PPP} $\Phi_\text{I}$ with density $\lambda_\text{I}$ \cite{HaeGan:09} on the ground. 
We then show how the analysis results for the single interfering node case in Section \ref{sec:analytical} can be used to obtain the outage probability for multiple interfering nodes case and the network outage probability.

When the locations of transmitting nodes are denoted by $u \in \Phi_\text{I}$,
a typical receiving \ac{UAV} will be associated with the nearest transmitting node $u_\text{o}$ and the other transmitting nodes become interfering nodes $u \in \Phi_\text{I} \backslash \{u_\text{o}\}$.\footnote{By Slivnyak's theorem \cite{StoKenMec:96}, we can obtain the network outage probability using the \ac{PPP} $\Phi_\text{I}$. Hence, $p_\text{o,m}^{\text{net}}$ is obtained using $p_\text{o,m}(\ell_\text{m})$ in \eqref{eq:multi}.}
In this network, the nearest transmittig node has the largest expected received power since the Tx with the smallest distance has the lowest path loss exponent, the largest Rician factor, and the highest \ac{LoS} probability \cite{KimLeeQue:19}.
Based on the association rule, the network outage probability can be obtained in the following corollary.

\begin{corollary} \label{col:multiple}
		When the typical receiving \ac{UAV} selects the nearest transmitting node,
		the network outage probability $p_\text{o,m}^\text{net}$ can be presented as
		\begin{align}\label{eq:PoutNet}
		p_\text{o,m}^{\text{net}}
		&=
		\mathbb{E} \left[ 
		\mathbb{P} \left[
		h_\text{m}  < \frac{\gamma_\text{t} I} {\Beta{m}}
		\bigg| I, \ell_\text{m} 
		\right]\right] \nonumber \\
		&=
		\int_{0}^{\infty} p_\text{o,m}\left(\sqrt{r^2 + z_\text{o}^2}\right) f_{\Dist{m}{H}}(r) \, dr
		\end{align}
		where $f_{\Dist{m}{H}}(r)=2 \lambda_\text{I} \pi r \exp(-\lambda_\text{I} \pi r^2)$ is the \ac{PDF} of the horizontal distance to the nearest node in a \ac{PPP} \cite{AndBacGan:11} and $\ell_\text{m}=\sqrt{r^2 + z_\text{o}^2}$ is the horizontal distance to the Tx $r$.
		In \eqref{eq:PoutNet}, the outage probability $p_\text{o,m}(\ell_\text{m})$ for the given link distance $\ell_\text{m}$ of the main link is presented as
		\begin{align}	
		&p_\text{o,m}(\ell_\text{m}) = \nonumber \\
		&\left\{
		1 -	\sum_{k=0}^{m-1} \frac{1}{k!} 
		\left(-\frac{m \gamma_\text{t}} {\Beta{m}}\right)^k
		\left[\frac{\partial} {\partial s^k} \mathcal{L}_I(s)\right]_
		{s=\frac{m \gamma_\text{t}} {\Beta{m}}}
		\right\}  \PLoSm
		\nonumber \\
		& +  \left[ 1 - \exp  \left\{ -2 \pi \lambda_\text{I}  
		\int_{r}^{\infty} \hspace{-1mm}
		\sum_{e_\text{I} \in \{\text{L},\text{N}\}} \hspace{-1mm}
		\left( 1 - \hat{p}_\text{o}^{(\text{N}, e_\text{I})} 
		\left(\sqrt{t^2 + z_\text{o}^2}\right)\right) \nonumber \right. \right.\\
		&\times \left. \left.
		p_{e_\text{I}}(t)  t \, dt 
		\right\} \right] 	\PNLoSm 
		\label{eq:multi} 
		\end{align}
		where $\hat{p}_\text{o}^{(e_\text{m}, e_\text{I})}
		\left(\sqrt{t^2 + z_\text{o}^2}\right)$ is the outage probability for an arbitrary interfering node in \eqref{eq:outage_h}
		and $\mathcal{L}_I(s)$ is the Laplace transform of the interference $I$, given by
		\begin{align} \label{laplace}
		\mathcal{L}_I(s)
		&  = 
		\exp \left\{ 
		-2 \pi \lambda_\text{I} \int_{r}^{\infty} \hspace{-1mm}
		\sum_{e_\text{I} \in \{\text{L},\text{N}\}} 
		\left( 1 -  \hat{p}_\text{o}^{(\text{L}, e_\text{I})}  
		\left(\sqrt{t^2 + z_\text{o}^2}\right) \hspace{-0.5mm} \right)
		\nonumber \right.\\
		& \quad \times \hspace{-1mm} \left.
		p_{e_\text{I}}(t)  t  \, dt 
		\right\} . 
		\end{align}
\end{corollary}
\begin{IEEEproof}
	See Appendix~\ref{app:multiple}.
\end{IEEEproof}

Using $p_\text{o,m}(\ell_\text{m})$ in \eqref{eq:multi}, 
we can also present the network outage probability, which is the average outage probability of links, distributed over the network.

From Corollary \ref{col:multiple}, we can see that the network outage probability are readily obtained using the outage probabilities with single interfering node, i.e., \eqref{eq:Po1h}, \eqref{eq:Po2h}, \eqref{eq:Po3h}, and \eqref{eq:Po4h}. 
Hence, the outage probability $\Pouth^{({e_\text{m},e_\text{I}})}(\mathcal{D})$ can be usefully used for various scenarios of \ac{UAV} communications for the performance analysis.

%
%
\section{Numerical Results}\label{sec:numerical}
In this section, we evaluate the outage probability of the \ac{UAV} communication and present the effects of the \ac{UAV} height, system parameters, and the channel environment on the outage probability.
We first compare the \ac{LoS} probabilities, which depend on the horizontal and vertical distances.
We then compare the general environment-based and the interference limited environment-based analysis results of outage probabilities, and then show the effects of \ac{UAV} height and the link environments on the outage probabilities.
We also show how the outage probability is changed for multiple interfering nodes case, compared to the case of considering one critical interfering node.
%
%
%
\begin{table}[!t]
	\caption{Parameter values if not otherwise specified \label{table:parameter}} 
	\begin{center}
		\rowcolors{2}
		{cyan!15!}{}
		\renewcommand{\arraystretch}{1.5}
		\begin{tabular}{l l | l l}
			\hline 
			{\bf Parameters} & {\bf Values} & {\bf Parameters} & {\hspace{0.32cm}}{\bf Values} \\
			\hline 
			\hspace{0.15cm}$\alpha_\text{N}$  & \hspace{0.2cm}$3.5$ 
			& \hspace{0.12cm}$\alpha_\text{L}$ & \hspace{0.2cm}$2$ \\ 
			\hspace{0.15cm}$\Pm$ [W] & \hspace{0.2cm}$10^{-8}$ 
			& \hspace{0.12cm}$N_\text{o}$ [W] & \hspace{0.2cm}$5\times10^{-17}$  \\ 
			\hspace{0.2cm}$K_\text{N}$  & \hspace{0.2cm}$1$ 
			& \hspace{0.2cm}$K_\text{L}$  & \hspace{0.2cm}$15$ \\ 
			\hspace{0.2cm}$\nu$  & \hspace{0.2cm}$3\times10^{-4}$ 
			& \hspace{0.12cm}$\mu$   & \hspace{0.2cm}$0.5$  \\ 
			\hspace{0.2cm}$\zeta$  & \hspace{0.2cm}$20$ 
			&\hspace{0.12cm}$W$ [Hz] & \hspace{0.2cm}$10^4$   \\
			\hspace{0.2cm}$\gamma_\text{t}$  & \hspace{0.2cm}$2$ 
			&\hspace{0.2cm}$ $  & \hspace{0.2cm}$ $\\ 
			\hline
		\end{tabular}
	\end{center}
\end{table}%
%
%
%
\begin{figure}[t!]
	\begin{center}   
		{ 
			\includegraphics[width=0.8\columnwidth]{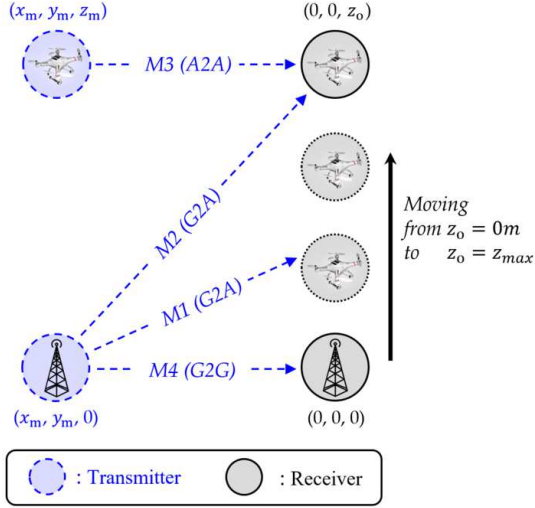}
			\vspace{-4mm}
		}
	\end{center}
	\caption{
		Simulation scenarios for the main links used in numerical results.
	}
	\label{fig:system1}
\end{figure}
%
%
\begin{figure}[t!]
	\begin{center}   
		{ 
			\includegraphics[width=0.9\columnwidth]{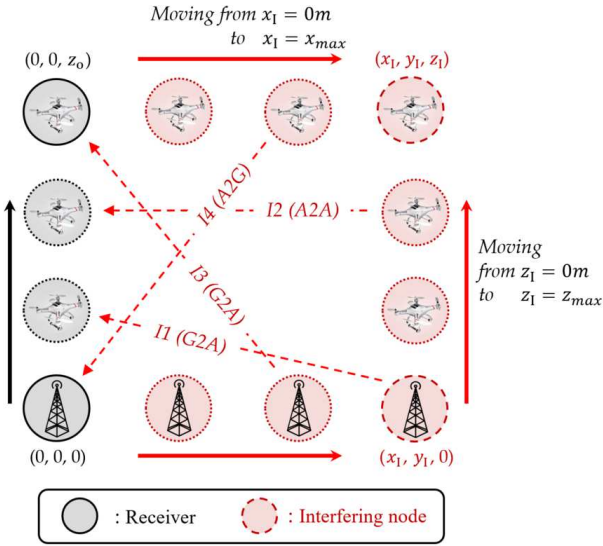}
			\vspace{-4mm}
		}
	\end{center}
	\caption{
		Simulation scenarios for the interference links used in numerical results.
	}
	\label{fig:system2}
\end{figure}
%

For convenience, we present the simulation scenarios
in Fig.~\ref{fig:system1} and Fig.~\ref{fig:system2}, where $M1-M4$ present the main link between a Tx and a Rx, while $I1-I4$ present the interference link between an interfering node and a Rx.
The solid-line arrows mean the case when the node moves in that direction.
Unless otherwise specified, the values of simulation parameters presented in Table \ref{table:parameter} are used. Note that the values of $\zeta$, $\nu$, and $\mu$ are adopted from \cite{BorElYan:16} for the dense urban environment.
%

%
\begin{figure}[t!]
	\begin{center}   
		{ 
			\psfrag{AAAAAAAAAAAAAAAAAAAAAAAAAA111}[Bl][Bl][0.59]{LoS probability in \cite{3GPP:TR:36.777:V15.0.0} with urban macro}
			\psfrag{AAAAAAAAAAAAAAAAAAAAAAAAAA12}[Bl][Bl][0.59]{LoS probability in \cite{3GPP:TR:36.777:V15.0.0} with urban micro}
			\psfrag{B1}[Bl][Bl][0.59]{LoS probability in \cite{BaiVazHea:14}}
			\psfrag{C1}[Bl][Bl][0.59]{LoS probability in \cite{HouKanLar:14}}
			\psfrag{X1}[Bl][Bl][0.59]{$\text{Height of UAV},\: d_\text{m}^{(\text{V})} \:[m]$}
			\psfrag{Y1}[Bl][Bl][0.59]{LoS Probability, $\PLoSm$}
			\includegraphics[width=1.00\columnwidth]{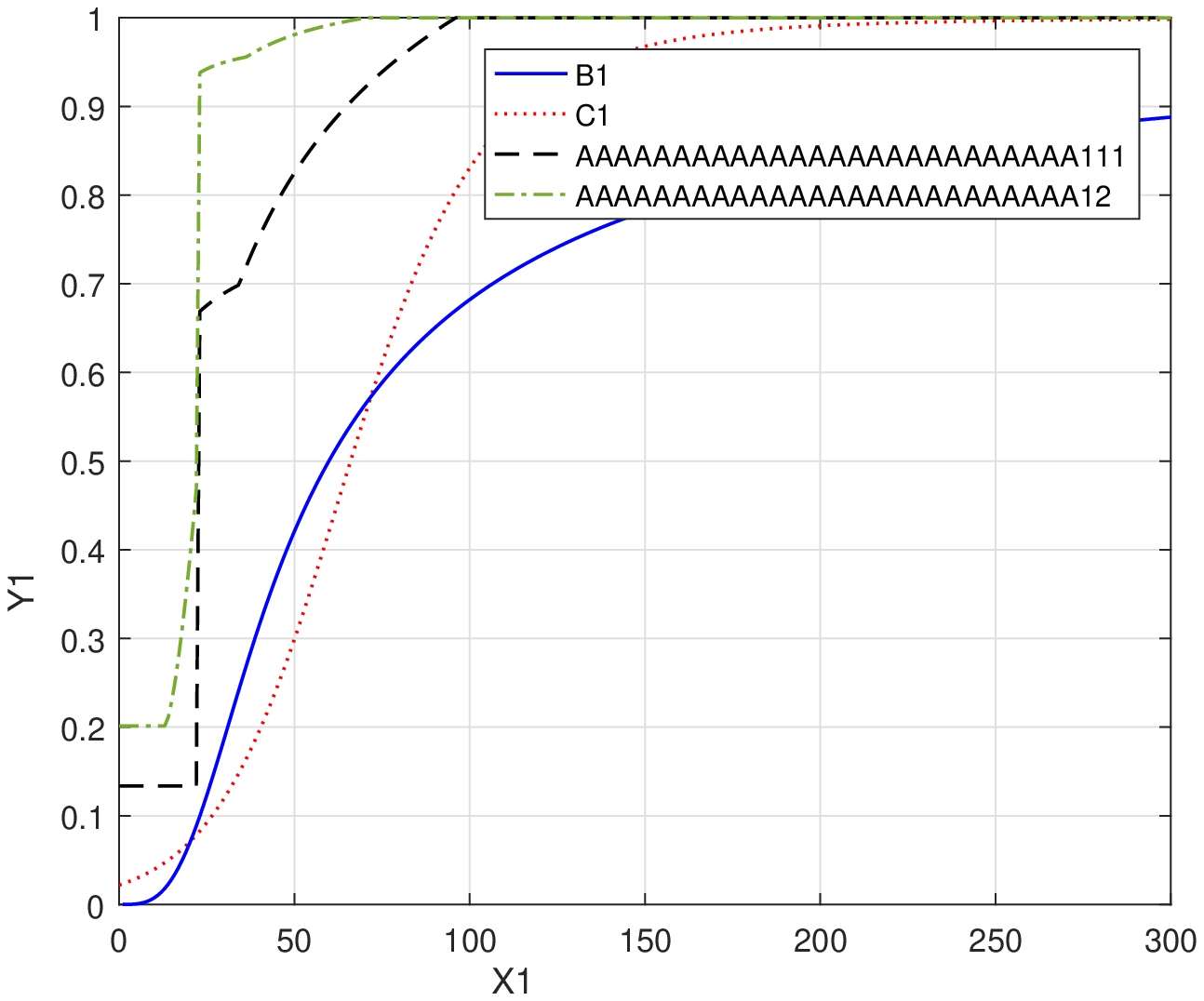}
			\vspace{-6mm}
		}
	\end{center}
	\caption{
		LoS probabilities $\PLoSm$ as a function of $d_\text{m}^{(\text{V})}$.}
	\label{fig:LoS_3GPP}
\end{figure}
%
%
\begin{figure}[t!]
	\begin{center}   
		{ 
			\psfrag{AAAAAAAAAAAAAAAAAAA}[Bl][Bl][0.59]{G2A channel $(0m-50m)$}
			\psfrag{BB}[Bl][Bl][0.59]{G2A channel $(0m-100m)$}
			\psfrag{CC}[Bl][Bl][0.59]{A2A channel $(25m-25m)$}
			\psfrag{DD}[Bl][Bl][0.59]{A2A channel $(25m-50m)$}
			\psfrag{EE}[Bl][Bl][0.59]{A2A channel $(50m-50m)$}
			\psfrag{Y}[Bl][Bl][0.59]{LoS Probability, $\PLoSm$}
			\psfrag{X}[Bl][Bl][0.59]{Horizontal distance of main link, $d_\text{m}^{(\text{H})} \:[m]$}
			\includegraphics[width=1.00\columnwidth]{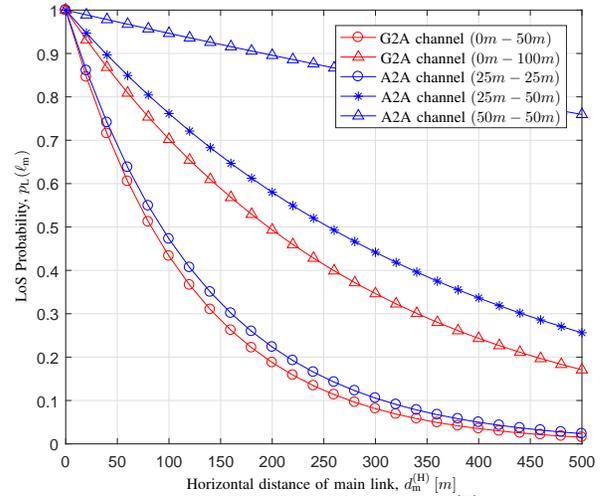}
			\vspace{-10mm}
		}
	\end{center}
	\caption{
		LoS probability $\PLoSm$ as a function of $d_\text{m}^{(\text{H})}$ with dense urban environments for different values of $z_\text{o}$ and $z_\text{m}$.
	}
	\label{fig:LoS}
\end{figure}
%
\subsection{Channel Components}
In this subsection, we first compare the \ac{LoS} probabilities in \cite{HouKanLar:14,BaiVazHea:14} and the \ac{LoS} probability of \ac{3GPP} model \cite{3GPP:TR:36.777:V15.0.0}, which are the most widely used for \ac{UAV} communication channels. We then also analyze the \ac{LoS} probabilities of the \ac{G2A} and \ac{A2A} channels.

	First, the \ac{LoS} probabilities in \cite{HouKanLar:14,BaiVazHea:14,3GPP:TR:36.777:V15.0.0} are compared in Figure 5, which shows $\PLoSm$ as a function of the \ac{UAV} height $\Dist{m}{V}$.
	The Tx is located at $(x_\text{m},y_\text{m},0)$, while the Rx moves from $(0,0,0)$ to $(0,0,z_\text{m})$ (i.e., $M1$ case with $\Dist{m}{H}=150m$).
From Fig.~\ref{fig:LoS_3GPP}, we can see that the \ac{LoS} probability of \ac{3GPP} model \cite{3GPP:TR:36.777:V15.0.0} has some limitaions.
Specifically, the \ac{LoS} probability of 3GPP model is constant regardless of the \ac{UAV} height, when the \ac{UAV} height is below 22.5$m$, and
it increases dramatically at around 22.5$m$ of the \ac{UAV} height, 
which might not be true in reality.
On the other hands, the \ac{LoS} probabilities in \cite{HouKanLar:14} and \cite{BaiVazHea:14} do not have above limitations, but the one in \cite{BaiVazHea:14} is only valid when the height of the ground device is much smaller than that of the \ac{UAV}.
Hence, we consider the \ac{LoS} probability model in \cite{BaiVazHea:14}.

Figure~\ref{fig:LoS} presents the \ac{LoS} probability $\PLoSm$ as a function of the horizontal distance of main link $\Dist{m}{H}$ for different values of $z_\text{o}$ and $z_\text{m}$. The Rx is located at $(0,0,z_\text{o})$ and the Tx moves from $(0,0,z_\text{m})$ to $(x_\text{m},y_\text{m},z_\text{m})$.
From this figure, we can see that the \ac{LoS} probability is a decreasing function as $\Dist{m}{H}$ increases because the elevation angle between a Tx and a Rx decreases with $\Dist{m}{H}$.
From Fig.~\ref{fig:LoS}, we can also see that the \ac{LoS} probability of the \ac{A2A} channel is generally higher than the that of the \ac{G2A} channel since the blockage effect by the obstacle reduces on the \ac{A2A} channel.
However, depending on the height difference between the Tx and the Rx, the \ac{LoS} probability of the \ac{G2A} channel (e.g., $z_\text{m}=0m$ and $z_\text{o}=100m$) can be higher than that of the \ac{A2A} channel (e.g., $z_\text{m}=25m$ and $z_\text{o}=25m$).
This is because the elevation angle of the \ac{G2A} channel is such large, so the probability of forming the \ac{LoS} link increases.

\subsection{General Environments vs. Interference-limited Environments}
\begin{figure}[t!]
	\begin{center}   
		{ 
			
			\psfrag{B}[Bl][Bl][0.59]{Outage Probability, $\Pout(\mathcal{D})$}
			\psfrag{A}[Bl][Bl][0.59]{$\text{Horizontal distance of interference link},\: d_\text{I}^{(\text{H})} \:[m]$}
			\psfrag{AAAAAAAAAAAAAAAAAAAAAAAAAAAAAAA}[Bl][Bl][0.59]{$\text{Interference-limited Environment (Analysis)}$}
			\psfrag{BB}[Bl][Bl][0.59]{$\text{Interference-limited Environment (Simulation)}$}
			\psfrag{C}[Bl][Bl][0.59]{$\text{General Environment (Analysis)}$}
			\psfrag{D}[Bl][Bl][0.59]{$\text{General Environment (Simulation)}$}
			\includegraphics[width=1.00\columnwidth]{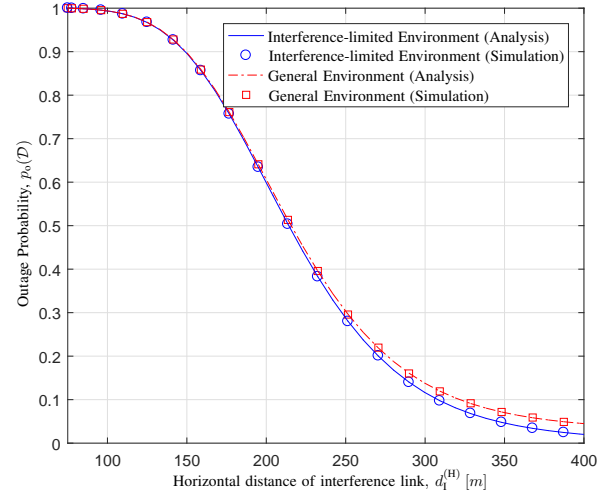}
			\vspace{-10mm}
		}
	\end{center}
	\caption{
		Outage probability $\Pout(\mathcal{D})$ as a function of $\Dist{I}{H}$ with $\PI=\Pm$, $\Dist{m}{H}=180m$, and $\Dist{m}{V}=\Dist{I}{V}=75m$.
	}
	\label{fig:simulation}
\end{figure}

Figure~\ref{fig:simulation} presents the outage probability $\Pout(\mathcal{D})$ as a function of the horizontal distance of the interference link $\Dist{I}{H}$,
where the Tx and the Rx are located at $(x_\text{m},y_\text{m},0)$ and $(0,0,z_\text{o})$, respectively (i.e., $M2$ case), while the interfering node moves from $(0,0,0)$ to $(x_\text{I},y_\text{I},0)$ (i.e., $I3$ case). Here, we use $\PI=\Pm$, $\Dist{m}{H}=180m$, and $\Dist{m}{V}=\Dist{I}{V}=75m$.
From this figure, we can first see that the analysis results closely match with the simulation results. In addition, the outage probability decreases as $\Dist{I}{V}$ increases.
This is because as $\Dist{I}{H}$ increases, the \ac{LoS} probability of the interference link decreases while the interference link distance increases with $\Dist{I}{H}$, which results in smaller interference at the Rx.
From Fig.~\ref{fig:simulation}, 
we can also see that the outage probability with the general environment (i.e., SINR-based case) has a similar trend to that with the interference-limited environment (i.e., SIR-based case). Hence, in the following figures, we present the numerical results of the interference-limited environments.
%
%
%
%
\begin{figure}[t!]
	\begin{center}   
		{ 
			\psfrag{A}[Bl][Bl][0.59]{Outage Probability, $\Pout(\mathcal{D})$}
			\psfrag{AAAAAAAAAAAAAAAAAAAAA}[Bl][Bl][0.59]{$\gamma_\text{t}=2,\: \ell_\text{I}=400m  \: \text{(G2A-A2A)}$}
			\psfrag{BBBBBBBBBBBBBBBBBBBBB}[Bl][Bl][0.59]{$\gamma_\text{t}=1,\: \ell_\text{I}=400m  \: \text{(G2A-A2A)}$}
			\psfrag{CCCCCCCCCCCCCCCCCCC}[Bl][Bl][0.59]{$\gamma_\text{t}=1,\: 
				\ell_\text{I}=450m  \: \text{(G2A-A2A)}$}
			\psfrag{DDDDDDDDDDDDDDDDDDDD}[Bl][Bl][0.59]{$\text{Simulation}$}
			\psfrag{D}[Bl][Bl][0.59]{$\PI=\Pm$}
			\psfrag{E}[Bl][Bl][0.59]{$\PI=0.3 \Pm$}
			\psfrag{Y}[Bl][Bl][0.59]{$\text{Height of UAV},\: \Dist{m}{V} \:[m]$}
			\includegraphics[width=1.00\columnwidth]{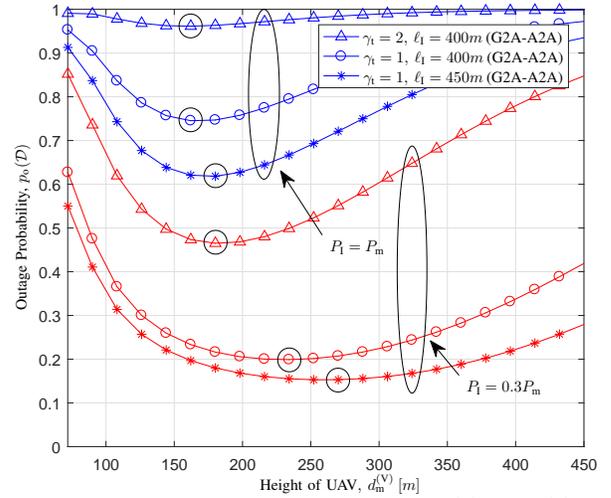}
			\vspace{-10mm}
		}
	\end{center}
	\caption{
		Outage probability $\Pout(\mathcal{D})$ as a function of $\Dist{m}{V}$ with $\Dist{m}{H}=80m$ for different values of $\gamma_\text{t}$, $\ell_\text{I}$, and $\PI$. The optimal UAV heights that minimize $\Pout(\mathcal{D})$ are marked by circles.
	}
	\label{fig:SIR1}
\end{figure}
%

\subsection{Effects of \ac{UAV} Height}
In this subsection, we show the impact of the \ac{UAV} height on the outage probability according to system parameters.

Figure \ref{fig:SIR1} presents the outage probability $\Pout(\mathcal{D})$ as a function of the \ac{UAV} height $\Dist{m}{V}$.
The Tx is located at $(x_\text{m}, y_\text{m}, 0)$, while the Rx and the interfering node move from $(0, 0, 0)$ to $(0, 0, z_\text{o})$ (i.e., $M1$ case) and move from $(x_\text{I}, y_\text{I}, 0)$ to $(x_\text{I}, y_\text{I}, z_\text{I})$ (i.e., $I2$ case), respectively.
Here, we use $\Dist{m}{H}=80m$ and different values of $\gamma_\text{t}$, $\ell_\text{I}$, and $\PI$.
To focus on the impact of the \ac{UAV} height on $\Pout(\mathcal{D})$, the environment of the interference link is set to be the same over different height of the \ac{UAV}, i.e., the interfering node is always located with the fixed distance $\ell_\text{I}$ to the Rx and has the \ac{A2A} channel.
%
From Fig.~\ref{fig:SIR1}, we can see that the outage probability first increases since the \ac{LoS} probaility of the interference link rapidly increases at a small height.
After the \ac{LoS} probability of the interference link increases to the end (i.e., $\PLoSi=1$), we can see that the outage probability first decreases when the \ac{UAV} height increases up to a certain value of the \ac{UAV} height, and then increases.
This is because the \ac{LoS} probability of the main link increases as the \ac{UAV} height increases.
For small UAV height, as the height increases, 
the increasing probability of forming \ac{LoS} main link affects more dominantly than the increasing main link distance on the outage probability.
However, for large \ac{UAV} height, the \ac{LoS} probability does not change that much with the height while the link distance becomes longer, so the outage probability increases.
We can also see that the optimal height above a certain \ac{UAV} height that minimizes $\Pout(\mathcal{D})$ decreases as the target SIR $\gamma_\text{t}$ or the power of the interfering node $\PI$ increases or the distance of the interference link $\ell_\text{I}$ decreases. 
From this, we can know that the optimal height decreases to reduce the main link distance as the impact of the interference link on the communication improves.
\begin{figure}[t!]
	\begin{center}   
		{ 
			\psfrag{A}[Bl][Bl][0.59]{Outage Probability, $\Pout(\mathcal{D})$}
			\psfrag{AAAAAAAAAAAAAAAAAAAAAA}[Bl][Bl][0.59]{$\gamma_\text{t}=3,\: d_\text{I}^{(\text{H})}=110m  \: \text{(G2A-G2A)}$}
			\psfrag{BBBBBBBBBBBBBBBBBBBBBB}[Bl][Bl][0.59]{$\gamma_\text{t}=2,\: d_\text{I}^{(\text{H})}=110m  \: \text{(G2A-G2A)}$}
			\psfrag{CCCCCCCCCCCCCCCCCCCCC}[Bl][Bl][0.59]{$\gamma_\text{t}=2,\: d_\text{I}^{(\text{H})}=130m  \: \text{(G2A-G2A)}$}
			\psfrag{D}[Bl][Bl][0.59]{$\PI=10 \Pm$}
			\psfrag{E}[Bl][Bl][0.59]{$\PI=\Pm$}
			\psfrag{B}[Bl][Bl][0.59]{Height of UAV, $\Dist{o}{V} \:[m]$ 
				$\left(\Dist{m}{V} =\Dist{I}{V} = \Dist{o}{V}\right)$}
			\includegraphics[width=1.00\columnwidth]{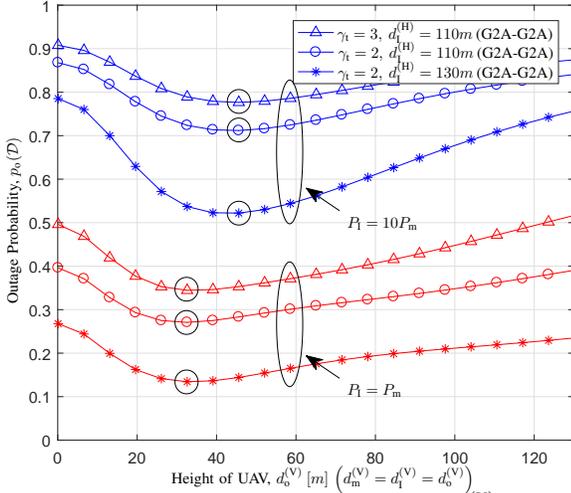}
			\vspace{-10mm}
		}
	\end{center}
	\caption{
		Outage probability $\Pout(\mathcal{D})$ as a function of $\Dist{o}{V}$ where $\Dist{m}{V} = \Dist{I}{V} = \Dist{o}{V}$ with $\Dist{m}{H}=80m$ for different values of $\gamma_\text{t}$, $\Dist{I}{H}$, and $P_\text{I}$. The optimal UAV heights that minimize $\Pout(\mathcal{D})$ are marked by circles.
	}
	\label{fig:SIR2}
\end{figure}
%

Figure \ref{fig:SIR2} presents the outage probability $\Pout(\mathcal{D})$ as a function of the \ac{UAV} height $\Dist{o}{V}$.
The Tx is located at $(x_\text{m}, y_\text{m}, 0)$ (i.e., $M1$ case) and the interfering node is located at $(x_\text{I}, y_\text{I}, 0)$ (i.e., $I1$ case), while the Rx moves from $(0, 0, 0)$ to $(0, 0, z_\text{o})$.
Here, we use $\Dist{m}{H}=80m$ and different values of $\gamma_\text{t}$, $\Dist{I}{H}$, and $\PI$.
To focus on the impact of the \ac{UAV} height on $\Pout(\mathcal{D})$, 
we vary the height of the Rx, i.e., $\Dist{o}{V}$, where $\Dist{m}{V} = \Dist{I}{V} = \Dist{o}{V}$,
and the Tx and the interfering node are located on the ground.
In this case, the \ac{LoS} probability of the main link is higher than that of the interference link due to $\Dist{m}{H} < \Dist{I}{H}$.
%
From Fig.~\ref{fig:SIR2}, we can see that the outage probability first decreases as the height increases up to a certain value of the \ac{UAV} height, and then increases.
This is because not only the \ac{LoS} probability of the main link but also that of the interference link increase with the \ac{UAV} height.
However, for large \ac{UAV} height, the \ac{LoS} probability of the interference link increases more than that of the main link.
We can also see that the optimal height increases as $\gamma_\text{t}$ or $\PI$ increases or $\ell_\text{I}$ decreases to improve the \ac{LoS} probability of the main link unlike the case in Fig.~\ref{fig:SIR1}.
\begin{figure}[t!]
	\begin{center}   
		{ 
			\psfrag{A}[Bl][Bl][0.59]{Outage Probability, $\Pout(\mathcal{D})$}
			\psfrag{AAAAAAAAAAAAAAAAA}[Bl][Bl][0.59]{$\Dist{I}{V}=100m \:\text{(G2G-A2G)}$}
			\psfrag{BBBBBBBBBBBBBBBBB}[Bl][Bl][0.59]{$\Dist{I}{V}=50m \: \text{(G2G-A2G)}$}
			\psfrag{CCCCCCCCCCCCCCCC}[Bl][Bl][0.59]{$\Dist{I}{V}=100m \: \text{(A2A-G2A)}$}
			\psfrag{DDDDDDDDDDDDDDDD}[Bl][Bl][0.59]{$\Dist{I}{V}=50m \: \text{(A2A-G2A)}$}
			\psfrag{AAAA}[Bl][Bl][0.59]{$\ell_{min}$}
			\psfrag{E}[Bl][Bl][0.59]{Horizontal distance of interference link, $\Dist{I}{H} \:[m]$}
			\includegraphics[width=1.00\columnwidth]{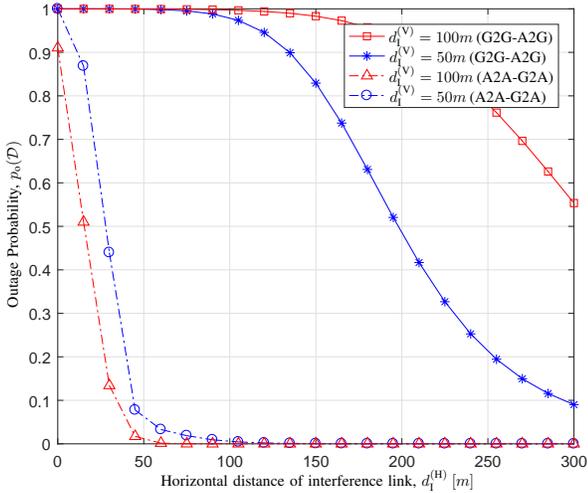}
			\vspace{-10mm}
		}
	\end{center}
	\caption{
		Outage probability $\Pout(\mathcal{D})$ as a function of $\Dist{I}{H}$ with $\PI=\Pm$ for different values of $\Dist{I}{V}$ and channel environment of the main link.
	}
	\label{fig:SIR3}
\end{figure}
%
\subsection{Effects of Main and Interference Link Environments}\label{sec:environment}
In this subsection, we focus on the impact of the environment of the main and interference links on the outage probability. 

Figure \ref{fig:SIR3} presents the outage probability $\Pout(\mathcal{D})$ as a function of the horizontal distance of the interference link $\Dist{I}{H}$ with $\PI=\Pm$ for different values of $\Dist{I}{V}$ and channel environment of the main link.
In Fig.~\ref{fig:SIR3}, two scenarios are considered: A2A main link with G2A interference link (A2A-G2A)
and G2G main link with A2G interference link (G2G-A2G).
The A2A-G2A case maps to $M3$ 
with $I3$, and the G2G-A2G case maps to $M4$ with $I4$ in Fig.~\ref{fig:system1} and Fig.~\ref{fig:system2}. 
Note that to explore the impact of the horizontal and vertical distances of interference link in this figure, 
the horizontal distance of interference link $\Dist{I}{H}$ is varied when the vertical distance $\Dist{I}{V} = 50m$ or $100m$.
To focus on the impact of the horizontal and vertical distance of the interference link, the main link is set as the \ac{A2A} or the \ac{G2G} channel with a fixed link distance 100$m$. 
The interference link is the \ac{A2G} or the \ac{G2A} channel. 
From this figure, we can see that generally, longer horizontal distance of the interference link (i.e., larger $\Dist{I}{H}$) results in lower outage probability. On the other hand, longer vertical distance of the interference link (i.e., larger $\Dist{I}{V}$) does not always result in upper outage probability.
Specifically, when the main link is the \ac{A2A} channel, the outage probability is smaller with $\Dist{I}{V}=100m$ than that with $\Dist{I}{V}=50m$. This is because, the \ac{LoS} probability of the main link with $\Dist{m}{V} = 50m$ is smaller than that with $\Dist{m}{V} = 100m$ even though the \ac{LoS} probability of the interference link with $\Dist{I}{V}=50m$ decreases faster than that with $\Dist{I}{V}=100m$ as $\Dist{I}{H}$ increases.

%
%
%
\begin{figure}[t!]
	\begin{center}   
		{ 
			\psfrag{B}[Bl][Bl][0.59]{$\Pout^{(\text{L,L})}(\mathcal{D})$, $\Pout^{(\text{N,N})}(\mathcal{D})$}
			\psfrag{AAAAAAAAAAAAAAAAAAAAAAAAAAAAAA}[Bl][Bl][0.59]{$\text{LoS main and interference links} \: \Pout^{(\text{L,L})}(\mathcal{D})$}
			\psfrag{BBBBBBBBBBBBBBBBBBBBBBBBBBBBBB}[Bl][Bl][0.59]{$\text{NLoS main and interference links} \: \Pout^{(\text{N,N})}(\mathcal{D})$}
			\psfrag{D}[Bl][Bl][0.59]{$\gamma_\text{t}=4$}
			\psfrag{E}[Bl][Bl][0.59]{$\gamma_\text{t}=2$}
			\psfrag{A}[Bl][Bl][0.59]{Ratio of Average received signal power, $\frac{\Beta{m}}  {\Beta{I}}$}
			\includegraphics[width=1\columnwidth]{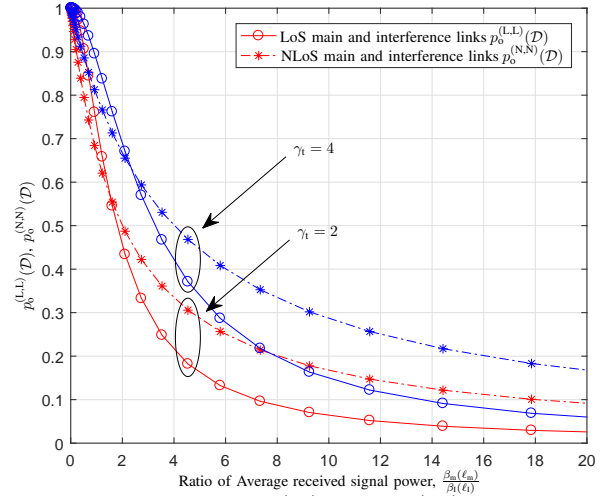}
			\vspace{-10mm}
		}
	\end{center}
	\caption{
		Outage probabilities $\Pout^{(\text{L,L})}(\mathcal{D})$ and $\Pout^{(\text{N,N})}(\mathcal{D})$ as a function of $\frac{\Beta{m}} {\Beta{I}}$ with $\Dist{I}{H}\hspace{-0.5mm}=\hspace{-0.5mm}100m$ and $\Dist{I}{V}\hspace{-0.5mm}=\hspace{-0.5mm}\Dist{m}{V}\hspace{-0.5mm}=\hspace{-0.5mm}70m$ for different values of $\gamma_\text{t}$.
	}
	\label{fig:SIRcompare}
\end{figure}
%
Figure \ref{fig:SIRcompare} presents the outage probabilities 
for LoS main and interference links $\Pout^{(\text{L,L})}(\mathcal{D})$ 
and NLoS main and interference links $\Pout^{(\text{N,N})}(\mathcal{D})$ as a function of $\frac{\Beta{m}}  {\Beta{I}}$ for different values of $\gamma_\text{t}$.
This is the case of $I3$ (G2A) with $M2$ (G2A) in Fig.~\ref{fig:system1} and Fig.~\ref{fig:system2}, and 
we use $\Dist{m}{H}=100m$ and $\Dist{I}{V}=\Dist{m}{V}=70m$.
From this figure, we can confirm that both outage probabilities are monotonic decreasing functions with 
$\frac{\Beta{m}}  {\Beta{I}}$. 
In addition, there exists a cross point of those probabilities at around $\frac{\Beta{m}}  {\Beta{I}} = 1.55$ when the target \ac{SIR} $\gamma_\text{t}=2$. For smaller  $\frac{\Beta{m}}  {\Beta{I}} < 1.55$,
$\Pout^{(\text{L,L})}(\mathcal{D})$ is greater than $\Pout^{(\text{N,N})}(\mathcal{D})$,
but it becomes opposite for larger $\frac{\Beta{m}}  {\Beta{I}} > 1.55$.
This verifies the results in Corollary~\ref{pro:outage compare} that the \ac{NLoS} environment can be more preferred for small $\frac{\Beta{m}}  {\Beta{I}}$.
We can also see that the value of the cross point increases from $1.55$ to $2.35$
as the target SIR $\gamma_\text{t}$ increases from $2$ to $4$.
Hence, we can know that the range of $\frac{\Beta{m}} {\Beta{I}}$ where the \ac{NLoS} environment is preferred increases as the target SIR $\gamma_\text{t}$ increases.

\subsection{Effects of Multiple Interfering Nodes}
In this subsection, we present how the outage probability is changed when we consider multiple interfering nodes, compared to the case of considering one dominant interfering node. 
Here, we define the dominant interfering node as the nearest one to the Rx, which gives 
the largest interference to the Rx on average.

When we consider one nearest interfering node among multiple interfering nodes, which are distributed in \ac{PPP} $\Phi_\text{I}$, the outage probability $p_\text{o,n}(\ell_\text{m})$ can also be obtained using the outage probability for single interfering node case $\Pouth^{({e_\text{m},e_\text{I}})}(\mathcal{D})$ in \eqref{eq:outage_h} as
\begin{align} \label{eq:nearest}
p_\text{o,n}(\ell_\text{m}) 
&= 
\int_{0}^{\infty} 
\sum_{e_\text{m}, e_\text{I} \in \{\text{L},\text{N}\}}^{} 
\Pouth^{(e_\text{m},e_\text{I})}  \left(\sqrt{r^2 + z_\text{o}^2}\right) \nonumber \\
&\quad\times 
p_{e_\text{m}}(\ell_\text{m}) p_{e_\text{I}}(r)  f_{\Dist{I}{H}}(r) \, dr 
\end{align}
where $\hat{p}_\text{o}^{(e_\text{m}, e_\text{I})}\left(\sqrt{r^2 + z_\text{o}^2}\right)$ is the outage probability for an arbitrary interfering node and $f_{\Dist{I}{H}}(r)=2 \lambda_\text{I} \pi r \exp(-\lambda_\text{I} \pi r^2)$ is the \ac{PDF} of the horizontal distance to the nearest interfering node from the Rx.
\begin{figure}[t!]
	\begin{center}   
		{ 
			\psfrag{A}[Bl][Bl][0.59]{\hspace{3mm}$p_\text{o,m}(\ell_\text{m})$, $p_\text{o,n}(\ell_\text{m})$}
			\psfrag{AAAAAAAAAAAAAAAAAAA}[Bl][Bl][0.59]{$\text{One nearest interfering node}$}
			\psfrag{BBBBBBBBBBBBBBBBBBB}[Bl][Bl][0.59]{$\text{Multiple interfering nodes}$}
			\psfrag{CC}[Bl][Bl][0.59]{$\gamma_\text{t}=2$}
			\psfrag{DD}[Bl][Bl][0.59]{$\gamma_\text{t}=3$}
			\psfrag{C}[Bl][Bl][0.59]{$\lambda_\text{I}=3\times10^{-5} [\text{nodes/}m^2]$}
			\psfrag{D}[Bl][Bl][0.59]{$\lambda_\text{I}=10^{-5} [\text{nodes/}m^2]$}
			\psfrag{E}[Bl][Bl][0.59]{$\lambda_\text{I}=2\times10^{-6} [\text{nodes/}m^2]$}
			\psfrag{B}[Bl][Bl][0.59]{Height of UAV, $\Dist{o}{V} \:[m]$ $\left(\Dist{m}{V} =\Dist{I}{V} = \Dist{o}{V}\right)$}
			\includegraphics[width=1.00\columnwidth]{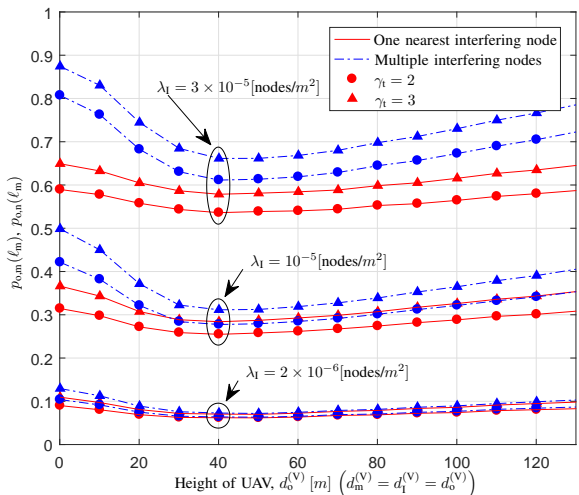}
			\vspace{-10mm}
		}
	\end{center}
	\caption{
		Outage probabilities with multiple interfering nodes $p_\text{o,m}(\ell_\text{m})$ and with one nearest interfering node $p_\text{o,n}(\ell_\text{m})$
		as a function of $\Dist{o}{V}$ where $\Dist{m}{V} \hspace{-0.5mm}=\hspace{-0.5mm}\Dist{I}{V}\hspace{-0.5mm} = \hspace{-0.5mm}\Dist{o}{V}$ with $\Dist{m}{H}\hspace{-0.5mm}=\hspace{-0.5mm}80m$ and $\PI\hspace{-0.5mm}=\hspace{-0.5mm}\Pm$ for different values of $\lambda_\text{I}$ and $\gamma_\text{t}$.
	}
	\label{fig:multiple}
\end{figure}
%

Figure \ref{fig:multiple} presents the outage probability with multiple interfering nodes, $p_\text{o,m}(\ell_\text{m})$ in Corollary \ref{col:multiple}, and that with one nearest interfering node, $p_\text{o,n}(\ell_\text{m})$ in \eqref{eq:nearest}
as a function of the \ac{UAV} height $\Dist{o}{V}$ for different values of the interfering node density $\lambda_\text{I}$ and the target \ac{SIR} $\gamma_\text{t}$.
For this figure, the Tx is located at $(80m, 0, 0)$, and
the location of Rx is changed from $(0, 0, 0)$ to $(0, 0, 250m)$ (i.e., $M1$ case).
The multiple interfering nodes are located on the ground (i.e., $I1$ case). 
Here, we also use $R=5000m$, $\Dist{m}{H}=80m$, and $\PI=\Pm$.
	In addition, since the multiple interfering nodes are randomly distributed in PPP, $\Dist{I}{H}$ becomes random, of which \ac{PDF} depends on the interfering node density $\lambda_\text{I}$.
As Fig.~\ref{fig:SIR2}, the outage probability first decreases as the height increases up to a certain value of the \ac{UAV} height, and then increases.

From this figure, we can see that the outage probability for the case of considering one dominant interfering node has the similar trend with that for the multiple interfering nodes case.
The difference in the outage probability for those two cases increase as the interfering node density $\lambda_\text{I}$ increases. This is because, as $\lambda_\text{I}$ increases, although the dominant interfering node can be located closer to Rx and generate larger interference, 
the amount of the interference from multiple interfering nodes increases more in the multiple interfering node case, which makes larger difference in the outage probabilities.

However, when the \ac{UAV} height is the optimal (like around 70m in Fig.~\ref{fig:multiple}) in terms of minimizing the outage probability,  
the outage probabilities of those two cases become almost the same. 
Therefore, from this result, we can see that the analysis result for the case of considering one interfering node, presented in this work, can also be usefully used for the optimal design of \ac{UAV} networks with multiple interfering nodes
such as the optimal \ac{UAV} height determination.
\begin{figure}[t!]
	\begin{center}   
		{ 
			\psfrag{AAAAAAAAAAAAAAAAAAA}[Bl][Bl][0.59]{$\text{One nearest interfering node}$}
			\psfrag{BBBBBBBBBBBBBBBBBBB}[Bl][Bl][0.59]{$\text{Multiple interfering nodes}$}
			\psfrag{L1}[Bl][Bl][0.59]{$\lambda_\text{I}=10^{-6} [\text{nodes/}m^2]$}
			\psfrag{L2}[Bl][Bl][0.59]{$\lambda_\text{I}=5\times10^{-7} [\text{nodes/}m^2]$}
			\psfrag{X}[Bl][Bl][0.59]{$\text{Height of UAV},\: \Dist{m}{V} \:[m]$}
			\psfrag{Y}[Bl][Bl][0.59]{\hspace{3mm}$p_\text{o,m}(\ell_\text{m})$, $p_\text{o,n}(\ell_\text{m})$}
			\includegraphics[width=1.00\columnwidth]{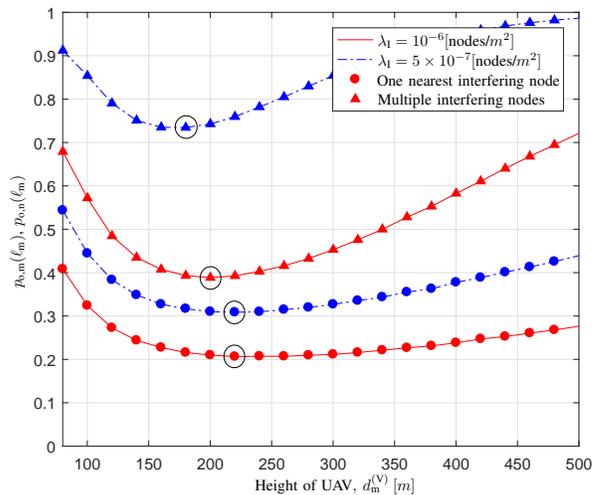}
			\vspace{-10mm}
		}
	\end{center}
	\caption{
		Outage probabilities with multiple interfering nodes $p_\text{o,m}(\ell_\text{m})$ and with one nearest interfering node $p_\text{o,n}(\ell_\text{m})$
		as a function of $\Dist{m}{V}$ with $d_\text{m}^{(\text{H})}=80m$ and $\PI=0.3\Pm$ for different values of $\lambda_\text{I}$.
	}
	\label{fig:multiple_A2A}
\end{figure}
%

Figure \ref{fig:multiple_A2A} presents the outage probability with multiple interfering nodes, $p_\text{o,m}(\ell_\text{m})$ in Corollary \ref{col:multiple}, and that with one nearest interfering node, $p_\text{o,n}(\ell_\text{m})$ in \eqref{eq:nearest}
as a function of the \ac{UAV} height $\Dist{m}{V}$ for different values of the interfering node density $\lambda_\text{I}$.
For this figure, the Tx is located at $(80m, 0, 0)$, and
the location of Rx is changed from $(0, 0, 0)$ to $(0, 0, 500m)$ (i.e., $M1$ case).
The multiple interfering nodes are located in the air (i.e., $I2$ case), and we use $R=5000m$, $\Dist{m}{H}=80m$, and $\PI=0.3\Pm$.

From this figure, we can see that the outage probability of one dominant interfering node has the similar trend with that of the multiple interfering nodes case.
However, since most of interfering nodes are in \ac{LoS} environment, the optimal \ac{UAV} heights that minimizes the outage probability of those two cases have a difference. 
Nevertheless, the optimal height of one dominant interfering node case can be used for the upper bound of that of the multiple interfering nodes case. Hence, we can see that the analysis result for the case of considering one interfering node, presented in this work, can also be used to give insights for the optimal design of \ac{UAV} networks with multiple interfering nodes even if the multiple interfering nodes are in the \ac{A2A} channel.
%
\begin{figure}[t!]
	\begin{center}   
		{ 
			\psfrag{A}[Bl][Bl][0.59]{\hspace{-7mm} Network Outage Probability, $p_\text{o,m}^\text{net}$}
			\psfrag{AAAAAAAAAAAAAAAAA}[Bl][Bl][0.59]{$\lambda_\text{I}=2\times10^{-6} [\text{nodes/}m^2]$}
			\psfrag{BB}[Bl][Bl][0.59]{$\lambda_\text{I}=10^{-5} [\text{nodes/}m^2]$}
			\psfrag{CC}[Bl][Bl][0.59]{$\lambda_\text{I}=5\times10^{-5} [\text{nodes/}m^2]$}
			\psfrag{B}[Bl][Bl][0.59]{Height of UAV, $\Dist{o}{V} \:[m]$ $\left(\Dist{m}{V} =\Dist{I}{V} = \Dist{o}{V}\right)$}
			\includegraphics[width=1.00\columnwidth]{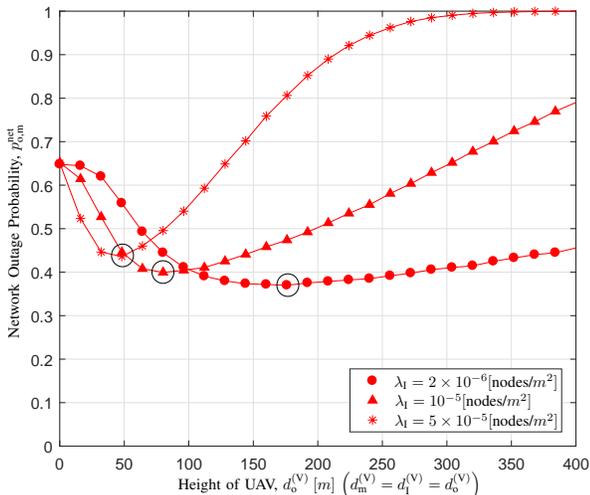}
			\vspace{-10mm}
		}
	\end{center}
	\caption{
		Network outage probability $p_\text{o,m}^\text{net}$
			as a function of $\Dist{o}{V}$ where $\Dist{m}{V} = \Dist{I}{V} = \Dist{o}{V}$ with $\PI=\Pm$ for different values of $\lambda_\text{I}$.
		}
		\label{fig:network}
	\end{figure}
	%

Figure~\ref{fig:network} shows the network outage probability as a function of UAV height
when $R=5000m$ and $\PI=\Pm$ for different values of the transmitting node density $\lambda_{\text{I}}$.
From this figure, we can see that as $\lambda_{\text{I}}$ increases, the optimal \ac{UAV} height decreases, while the optimal outage probability increases.
Lowering the optimal \ac{UAV} height can increase both the received interference power from other transmitting nodes and the main link received power. Hence, from the results of this figure, we can see that when $\lambda_{\text{I}}$ is larger, the optimal \ac{UAV} height becomes smaller as increasing the received power of the main link becomes more dominantly determined the outage probability than the increasing interference power.
%
%
%

\section{Conclusion}\label{sec:conclusion}
This paper analyzes the impact of the interfering node for reliable \ac{UAV} communications. 
After characterizing the channel model affected by the horizontal distance and the vertical distance of the communication link, 
we derive the outage probability in a closed-form for all possible scenarios of main and interference links. 
Furthermore, we show the effects of the transmission power, the horizontal link and vertical link distances, and  
the communication scenarios of main and interference links. 
Specifically, we show the existence of the optimal heights of the \ac{UAV} for various scenarios, which increase as the power of the interfering node decreases or the interference link distance increases. 
We also analytically prove that the \ac{NLoS} environment can be better than the \ac{LoS} environment if the average received power of the interference is much larger than that of the main link signal.
The outcomes of this work can be usefully used for the optimal height determination of \acp{UAV} in the presence of an interfering node, and it can give insights on the \ac{UAV} height for the multiple interfering nodes case as well.
%

\begin{appendix}
	\subsection{Proof of Theorem~\ref{trm:OPSINR}} \label{app:trm}

		As the main and interference links can be in either the \ac{LoS} or \ac{NLoS} environments when the probability is $\PLoS$ or $\PNLoS$, respectively, the outage probability is divided into four cases, which are $\Pout\!^{(\text{L,L})}(\mathcal{D})$, $\Pout\!^{(\text{L,N})}(\mathcal{D})$, $\Pout\!^{(\text{N,L})}(\mathcal{D})$, and $\Pout\!^{(\text{N,N})}(\mathcal{D})$ according to the environments of main and interference links.
		Hence, the outage probability is obtained as \eqref{eq:outage} using the law of total probability.
		We derive $\Pout^{({e_\text{m},e_\text{I}})}(\mathcal{D})$ for the above four cases as follows.

		For \emph{Case} 1, $\Km(\ell_\text{m}) \neq 0$ and $\Ki(\ell_\text{I}) \neq 0$ as both main and interference links are in \ac{LoS} environments, and $\Pout\!^{(\text{L,L})}(\mathcal{D})$ can be obtained using \eqref{eq:channel fading1} as
		\begin{align}
		\Pout\!^{(\text{L,L})}(\mathcal{D})
		=\int_{0}^{\infty} \hspace{-2mm}
		\int_{0}^{\frac{\gamma_\text{t} ( \Beta{I} g + N_\text{o}) }  {\Beta{m}}} 
		%
		%
		f_{h_\text{m}}(h) \, dh  f_{h_\text{I}}(g)  \, dg. \label{eq:pout1}
		\end{align}
		By using the \ac{CDF} of the noncentral Chi-squared distribution in \eqref{eq:pout1}, 
		$\Pout\!^{(\text{L,L})}(\mathcal{D})$ is presented as \eqref{eq:Po1}.\par
		In \emph{Case} 2, $\Km(\ell_\text{m})\neq 0$ and $\Ki(\ell_\text{I})= 0$ as the interference link is in the \ac{NLoS} environment, and $\Pout\!^{(\text{L,N})}(\mathcal{D})$ is obtained using \eqref{eq:channel fading1} and \eqref{eq:channel fading2} as 
		\begin{align}
		&\Pout\!^{(\text{L,N})}(\mathcal{D}) = 
		\int_{0}^{\infty} 
		\int_{0}^{\frac{\gamma_\text{t} ( \Beta{I} g + N_\text{o})}  {\Beta{m}}}  
		%
		%
		f_{h_\text{m}}(h) \, dh  f_{h_\text{I}}(g)  \, dg	 \nonumber \\
		&\overset{\underset{\mathrm{(a)}}{}}{=}
		%
		%
		1 - \hspace{-1mm} \int_{0}^{\infty}   \hspace{-1mm}
		Q \hspace{-0.5mm} \left(\hspace{-1mm} \sqrt{2 \Km(\ell_m)} , 
		\sqrt{\frac{ \gamma_\text{t} (\Beta{I} g \hspace{-0.5mm} 
				+ \hspace{-1mm} N_\text{o})}  {\Beta{m}}}\right) \hspace{-0.5mm}
		\exp \hspace{-0.5mm} \left(-g \right) \, \hspace{-1mm} dg \nonumber \\
		&\overset{\underset{\mathrm{(b)}}{}}{=}
		%
		%
		1 - \frac{\Beta{m}}  {\gamma_\text{t} \Beta{I}}
		\exp\left(\frac{N_\text{o}}  {\Beta{I}}\right) \nonumber \\
		&\quad \times \hspace{-1mm}
		\int_{\frac{\gamma_\text{t} N_\text{o}}  {\Beta{m}}} ^ {\infty} \hspace{-1mm}
		Q \hspace{-0.5mm} \left(\hspace{-0.5mm}\sqrt{2 \Km(\ell_\text{m})} , \sqrt{g'}\right)  
		\exp \hspace{-0.5mm} \left(\hspace{-0.5mm} - \frac{\Beta{m} g'}  
		{ \gamma_\text{t} \Beta{I}}\right) \,\hspace{-1mm}  dg' \hspace{-2mm} \label{eq:pout2}
		\end{align}
		where $Q(a,b)$ is the first-order Marcum Q-function.
		In \eqref{eq:pout2}, (a) is from the CDF of the noncentral Chi-squared distribution,
		(b) is obtained by substitution from $\frac{\gamma_\text{t} \Beta{I}} {\Beta{m}} g + \frac{\gamma_\text{t} N_\text{o}}  {\Beta{m}}$ to $g'$,
		and the integral term can be represented as
		\begin{align} \label{eq:marcum1}
		&\int_{\frac{d^2}{2}}^{\infty}  
		\exp\left(-c^2  x\right)   Q\left(e , f\sqrt{2x}\right) \, dx \nonumber \\
		&=
		\frac{1}  {c^2} \left\{\exp\left(-\frac{c^2 d^2}  {2}\right) Q(e,df) 
		- \frac{c^2}  {c^2 + f^2} \nonumber 
		\right.\\
		&\quad\left.
		\times \exp \hspace{-0.5mm}\left(\hspace{-0.5mm}-\frac{c^2 e^2}  {2(c^2 + f^2)}\hspace{-0.5mm}\right) \hspace{-0.5mm}
		Q\hspace{-0.5mm}\left(\hspace{-0.5mm}\frac{ef}  {\sqrt{c^2 + f^2}} , d\sqrt{c^2 + f^2}\right)\hspace{-1mm}\right\} \hspace{-1mm}
		\end{align}
		where $c=\sqrt{\frac{\Beta{m}} {\gamma_\text{t} \Beta{I}}}$, $d=\sqrt{\frac{2\gamma_\text{t} N_\text{o}} {\Beta{m}}}$, $e=\sqrt{2\Km(\ell_\text{m})}$, and $f=\sqrt{\frac{1}{2}}$ from \cite[eq. (40)]{Nut:72}.
		By using \eqref{eq:marcum1} in \eqref{eq:pout2}, 
		$\Pout\!^{(\text{L,N})}(\mathcal{D})$ is presented as \eqref{eq:Po2}.\par
		In \emph{Case} 3, $\Km(\ell_\text{m})= 0$ and $\Ki(\ell_\text{I})\neq 0$ as the main link is in the \ac{NLoS} environment, and $\Pout\!^{(\text{N,L})}(\mathcal{D})$ is given by
		\begin{align} \label{eq:pout3}
		&\Pout\!^{(\text{N,L})}(\mathcal{D}) = 
		\int_{0}^{\infty} 
		\int_{0}^{\frac{\gamma_\text{t} (\Beta{I} g + N_\text{o})}  {\Beta{m}}}  
		%
		%
		f_{h_\text{m}}(h) \, dh  f_{h_\text{I}}(g)  \, dg	 \nonumber \\
		&\overset{\underset{\mathrm{(a)}}{}}{=}
		%
		%
		1 - \frac{1}  {2}  \int_{0}^{\infty}  
		\exp\left(-\frac{\gamma_\text{t}(\Beta{I} g + N_\text{o})} {\Beta{m}}\right)  \nonumber \\
		&\quad\times \exp\left( - \Ki(\ell_\text{I}) - \frac{g}  {2} \right) 
		I_0 \left(\sqrt{2 \Ki(\ell_\text{I}) g}\right) \, dg.
		\end{align}
		In \eqref{eq:pout3}, (a) is from the CDF of the exponential distribution and 
		the integral term can be presented as
		\begin{align} \label{eq:bessel1}
		\int_{0}^{\infty}  \exp(-c^2 x)  I_0\left(d\sqrt{2x}\right)\,dx
		= 
		\frac{1}  {c^2}  \exp\left(\frac{d^2}  {2c^2}\right)  
		\end{align}
		where $c=\sqrt{\frac{1}{2} + \frac{\gamma_\text{t} \Beta{I}} {\Beta{m}}}$ and $d=\sqrt{K_\text{I}(\ell_\text{I})}$ from \cite[eq. (9)]{Nut:72}.
		By using \eqref{eq:bessel1} in \eqref{eq:pout3}, 
		$\Pout\!^{(\text{N,L})}(\mathcal{D})$ is presented as \eqref{eq:Po3}.\par
		In \emph{Case} 4, $\Km(\ell_\text{m})= 0$ and $\Ki(\ell_\text{I})= 0$ as the main and the interference links are both in \ac{NLoS} environments, and $\Pout\!^{(\text{N,N})}(\mathcal{D})$ is given by
		\begin{align} \label{eq:pout4}
		&\Pout\!^{(\text{N,N})}(\mathcal{D}) =
		\int_{0}^{\infty} 
		\int_{0}^{\frac{\gamma_\text{t} (\Beta{I} g + N_\text{o})}  {\beta_\text{m}(\ell_\text{m})}}  
		%
		%
		f_{h_\text{m}}(h) \, dh  f_{h_\text{I}}(g)  \, dg	 \nonumber \\
		&\overset{\underset{\mathrm{(a)}}{}}{=}
		%
		%
		1 - \int_{0}^{\infty} 
		\exp\left(-\frac{\gamma_\text{t}(\Beta{I} g +  N_\text{o})} {\Beta{m}} 
		- g  \right)  \, dg
		\end{align}
		where (a) is from the CDF of the exponential distribution.
		By simple calculation, $\Pout\!^{(\text{N,N})}(\mathcal{D})$ is presented as \eqref{eq:Po4}.
	\subsection{Proof of Lemma~\ref{lem:OPSIR}} \label{app:lem}
		In the interference-limited environment, the interfering signal power is much stronger than the noise power (i.e., $\Hi \Beta{I} \gg N_\text{o}$), so the noise is negligible.
		Consequently, the communication performance can be analyzed based on
		$\hat{\gamma}(\ell_\text{m},\ell_\text{I})
		=\frac{\Hm \Beta{m}}  {\Hi \Beta{I}}$
		instead of
		$\gamma(\ell_\text{m},\ell_\text{I})
		= \frac{\Hm \Beta{m}}  {\Hi \Beta{I} + N_\text{o}}$.
		Hence, the integral interval in the outage probability substitutes from $\left[0,\frac{\gamma_\text{t} (\Hi \Beta{I} + N_\text{o})}  {\Beta{m}}\right]$
		to $\left[0,\frac{\gamma_\text{t} \Hi \Beta{I}}  {\Beta{m}} \right]$, and
		we obtain $\Pouth^{({e_\text{m},e_\text{I}})}(\mathcal{D})$ for the above four cases as follows.

		For \emph{Case} 1, i.e., $\Km(\ell_\text{m}) \neq 0$ and $\Ki(\ell_\text{I}) \neq 0$, we can present $\Pouth\!^{(\text{L,L})}(\mathcal{D})$ by replacing $N_\text{o}=0$ in \eqref{eq:Po1} as
		\begin{align}
		\Pouth\!^{(\text{L,L})}(\mathcal{D})
		&=
		\int_{0}^{\infty} \hspace{-2mm}
		\int_{0}^{\frac{\gamma_\text{t} \Beta{I} g }  {\Beta{m}}} 
		%
		%
		f_{h_\text{m}}(h) \, dh  f_{h_\text{I}}(g)  \, dg \nonumber \\
		&\overset{\underset{\mathrm{(a)}}{}}{=}
		1 - \frac{1}  {2}  \int_{0}^{\infty}  
		Q\left(\sqrt{2 \Km(\ell_\text{m})}  ,
		\sqrt{\frac{\gamma_\text{t} \Beta{I} g}  {\Beta{m}}}\right)  \nonumber \\
		&\quad\times 
		\exp\left( - \Ki(\ell_\text{I})  -  \frac{g}  {2} \right)
		I_0\left(\sqrt{2 \Ki(\ell_\text{I}) g}\right) \, dg \label{eq:pout5}
		\end{align}
		where (a) is from the CDF of the noncentral Chi-squared distribution and the integral term in \eqref{eq:pout5} can be presented as
		\begin{align}  \label{eq:marcum2}
		&\int_{0}^{\infty}  \exp\left(-c^2 x\right)  I_0\left(d \sqrt{2x}\right)  
		Q\left(e , f \sqrt{2x}\right)\,dx  \nonumber \\
		&=
		\frac{1}  {c^2}  
		\left\{
		\exp\left(\frac{d^2}  {2c^2}\right)  
		Q\left(\frac{c e}  {\sqrt{{c}^2 + {f}^2}},  
		\frac{d f}  {c \sqrt{{c}^2 + {f}^2}}\right)  \nonumber \right.\\
		&\quad\left.
		- \frac{f^2}  {c^2 + f^2}
		\exp\left(\frac{d^2 - c^2 e^2}  {2(c^2 + f^2)}\right)
		I_0\left(\frac{d e f}  {c^2 + f^2}\right)
		\right\}
		\end{align}
		where $c =\sqrt{0.5}$, $d =\sqrt{\Ki(\ell_\text{I})}$, $e =\sqrt{2\Km(\ell_\text{m})}$, and $f=\sqrt{\frac{\gamma_\text{t} \Beta{I}} 
			{2 \Beta{m}}}$ from \cite[eq. (46)]{Nut:72}.
		By using \eqref{eq:marcum2} in \eqref{eq:pout5}, $\Pouth\!^{(\text{L,L})}(\mathcal{D})$ is presented as \eqref{eq:Po1h}.\par
		In \emph{Case} 2, i.e., $\Km(\ell_\text{m})\neq 0$ and $\Ki(\ell_\text{I})= 0$, $\Pouth\!^{(\text{L,N})}(\mathcal{D})$ is presented using \eqref{eq:pout2} as
		\begin{align} \label{eq:pout6}
		\Pouth\!^{(\text{L,N})}(\mathcal{D}) 
		&=
		\int_{0}^{\infty} \hspace{-2mm}
		\int_{0}^{\frac{\gamma_\text{t} \Beta{I} g }  {\Beta{m}}} 
		%
		%
		f_{h_\text{m}}(h) \, dh  f_{h_\text{I}}(g)  \, dg \nonumber \\
		&\overset{\underset{\mathrm{(a)}}{}}{=}
		%
		%
		1  - 
		\int_{0}^{\infty} 
		Q\left(
		\sqrt{2 \Km(\ell_\text{m})},
		\sqrt{\frac{\gamma_\text{t} \Beta{I} g}  {\Beta{m}}}\right) 
		\nonumber \\
		& \quad \times
		\exp\left(-g\right) \, dg
		\end{align}
		where (a) is from the CDF of the noncentral Chi-squared distribution and the integral term in \eqref{eq:pout6} can be presented as \eqref{eq:marcum1} with
		$c= 1$, $d= 0$ $e = \sqrt{2\Km(\ell_\text{m})}$, and $f =\sqrt{\frac{\gamma_\text{t} \Beta{I}} {2 \Beta{m}}}$.
		By using \eqref{eq:marcum1} in \eqref{eq:pout6}, $\Pouth\!^{(\text{L,N})}(\mathcal{D})$ is presented as \eqref{eq:Po2h}.\par
		In \emph{Case} 3, i.e., $\Km(\ell_\text{m})= 0$ and $\Ki(\ell_\text{I})\neq 0$, $\Pouth\!^{(\text{N,L})}(\mathcal{D})$ is obtained by making $N_\text{o}=0$ in \eqref{eq:Po3} as \eqref{eq:Po3h}.
		In \emph{Case} 4, i.e., $\Km(\ell_\text{m})= 0$ and $\Ki(\ell_\text{I})= 0$, $\Pouth\!^{(\text{N,N})}(\mathcal{D})$ is obtained by making $N_\text{o}=0$ in \eqref{eq:Po4} as \eqref{eq:Po4h}.
	\subsection{Proof of Corollary~\ref{col:multiple}} \label{app:multiple}
		For the multiple interfering nodes case, the outage probability can be presented as
		\begin{align}\label{eq:Pout}
		p_\text{o,m}(\ell_\text{m})&=
		p_\text{o,m}^{(\text{L})}(\ell_\text{m})  \PLoSm
		\hspace{-0.5mm} + \hspace{-0.5mm}
		p_\text{o,m}^{(\text{N})}(\ell_\text{m}) \PNLoSm  \hspace{-1mm}
		\end{align}
		where $p_\text{o,m}^{(\text{L})}(\ell_\text{m})$
		and $p_\text{o,m}^{(\text{N})}(\ell_\text{m})$
		are the outage probabilities for \ac{LoS} and \ac{NLoS} main links, respectively.
		When we consider the Nakagami-m fading for the \ac{LoS} link and the interference-limited environment to derive the outage probability tractably, $p_\text{o,m}^{(\text{L})}(\ell_\text{m})$ is given by
		\begin{align} \label{eq:multiple_L}
		&p_\text{o,m}^{(\text{L})}(\ell_\text{m}) =  
		\mathbb{E} \left[
		\mathbb{P}\left[ \Hm  < \frac{\gamma_\text{t} I} {\Beta{m}} \bigg| I
		\right] \right]
		\nonumber \\
		&  \overset{\underset{\mathrm{(a)}}{}}{=} 
		1 - \mathbb{E} 
		\left[ 
		\frac{\Gamma\left(m,\frac{m\gamma_\text{t} I} {\Beta{m}}\right)} {\Gamma(m)} 
		\right] \nonumber \\
		& \overset{\underset{\mathrm{(b)}}{}}{=} 
		1 - \mathbb{E}
		\left[
		\sum_{k=0}^{m-1} \frac{1}{k!} \left(\frac{m \gamma_\text{t} I} {\Beta{m}}\right)^k 
		\exp \left(-\frac{m \gamma_\text{t} I} {\Beta{m}}\right) \right] \nonumber \\
		&=
		1  - \sum_{k=0}^{m-1} 
		\frac{1}{k!} \left(-\frac{m \gamma_\text{t}} {\Beta{m}}\right)^k 
		\left[\frac{\partial} {\partial s^k} \mathcal{L}_I(s)\right]_
		{s=\frac{m \gamma_\text{t}} {\Beta{m}}}
		\end{align}
		where $I=\sum_{u \in \Phi_\text{I} \backslash \{u_\text{o}\}}^{}h_u \ell_u^{-\alpha_u(\ell_u)} \PI$ is the interference from multiple interfering nodes, (a) is obtained because $h_m\sim\Gamma(m,1/m)$, and (b) follows from the definition of incomplete gamma function for integer values of $m$.\footnote{Note that (a) can be obtained only when $m= \frac{\Km(\ell_\text{m})^2 + 2\Km(\ell_\text{m}) + 1} {2 \Km(\ell_\text{m}) + 1}$ is an integer. Hence, we cannot obtain $p_\text{o,m}^{(\text{L})}(\ell_\text{m})$ for all scenarios.} In \eqref{eq:multiple_L}, $\mathcal{L}_I(s)$ is the Laplace transform of the interference $I$, and is given by
		\begin{align}
		\mathcal{L}_I(s)
		& = 
		\mathbb{E}_{\Phi_\text{I} \backslash \{u_\text{o}\}} 
		\hspace{-1mm} \left[\exp\left(-s \sum_{u \in \Phi_\text{I} \backslash \{u_\text{o}\}}^{} P_\text{I} h_u \ell_u^{-\alpha_u(\ell_u)}\right) \right] \nonumber \\
		& = 
		\mathbb{E}_{\Phi_\text{I} \backslash \{u_\text{o}\}} 
		\hspace{-1mm} \left[	\prod_{u \in \Phi_\text{I} \backslash \{u_\text{o}\}}^{} \mathbb{E}_{h_u} \hspace{-0.5mm} 
		\left[	\exp\left(-s h_u \ell_u^{-\alpha_u(\ell_u)} \PI\right) 
		\right] \right] \nonumber \\
		& \overset{\underset{\mathrm{(a)}}{}}{=} 
		\exp \left\{ 
		-2 \pi \lambda_\text{I} \int_{r}^{\infty} 
		\sum_{e_\text{I} \in \{\text{L},\text{N}\}} 
		\left( 1 \hspace{-0.5mm} - \hspace{-0.5mm} \hat{p}_\text{o}^{(\text{L}, e_\text{I})} \hspace{-1mm} \left(\sqrt{t^2 + z_\text{o}^2}\right) \hspace{-0.5mm} \right)
		\nonumber \right.\\
		& \quad \times \hspace{-1mm} \left.
		p_{e_\text{I}}(t)  t  \, dt 
		\right\}
		\end{align}
		where (a) is from the \ac{PGFL} \cite{HaeGan:09}.
		In \eqref{eq:Pout}, $p_\text{o,m}^{(\text{N})}(\ell_\text{m})$ is given by
		\begin{align} \label{eq:multiple_N}
		&p_\text{o,m}^{(\text{N})}(\ell_\text{m}) = 
		\mathbb{E} \left[
		\mathbb{P}\left[ \Hm  < \frac{\gamma_\text{t} I} {\Beta{m}} \bigg| I
		\right] \right]
		\nonumber \\
		&\overset{\underset{\mathrm{(a)}}{}}{=}
		1 - \mathbb{E} \left[ 
		\exp\left(-\frac{\gamma_\text{t} I} {\Beta{m}}\right) 
		\right] \nonumber \\
		&=
		1 - \mathbb{E}_{\Phi_\text{I} \backslash \{u_\text{o}\}} 
		\hspace{-0.5mm} \left[	\prod_{u \in \Phi_\text{I} \backslash \{u_\text{o}\}}^{} \hspace{-0.5mm} \mathbb{E}_{h_u} \hspace{-0.5mm} 
		\left[\exp\hspace{-0.5mm} 
		\left(-\frac{\gamma_\text{t} h_u \ell_u^{-\alpha_u(\ell_u)} \PI} {\Beta{m}}\right) 
		\right] \right] \nonumber \\
		&\overset{\underset{\mathrm{(b)}}{}}{=}
		1   -   \exp  \left\{
		-2 \pi \lambda_\text{I}  \int_{r}^{\infty} 
		\sum_{e_\text{I} \in \{\text{L},\text{N}\}} 
		\left( 1 - \hat{p}_\text{o}^{(\text{N}, e_\text{I})} 
		\left(\sqrt{t^2 + z_\text{o}^2}\right)\right) \nonumber \right.\\
		&\quad \times \left.
		p_{e_\text{I}}(t) t \, dt 
		\right\}
		\end{align}
		where (a) is obtained because $h_\text{m}\sim\exp(1)$ and (b) is from the \ac{PGFL}.
\end{appendix}

%

\bibliographystyle{IEEEtran}

\bibliography{StringDefinitions,IEEEabrv,mybib}

\end{document}